\documentclass[12pt]{article}
\usepackage{amssymb}
\usepackage{amsfonts}
\usepackage{amsmath}
\allowdisplaybreaks

\begin{document}

\author{C. Bizdadea\thanks{%
E-mail address: bizdadea@central.ucv.ro} $^{,1}$, S.\,O. Saliu\thanks{%
E-mail address: osaliu@central.ucv.ro}$\,\,^{,1}$\\
$^{1}$Faculty of Physics, University of Craiova,\\
13 Al. I. Cuza Str., Craiova 200585, Romania\\
E.\,M. B\u{a}b\u{a}l\^{\i}c\thanks{%
E-mail address: mbabalic@central.ucv.ro}$\,\,^{,1,2}$\\
$^{2}$Department of Theoretical Physics,\\
Horia Hulubei National Institute\\
of Physics and Nuclear Engineering,\\
PO Box MG-6, Bucharest, Magurele 077125, Romania}
\title{Selfinteractions in collections of \\
massless tensor fields \\
with the mixed symmetry $\left( 3,1\right) $ and $\left( 2,2\right) $}
\date{}
\maketitle

\begin{abstract}
Under the hypotheses of analyticity, locality, Lorentz covariance, and
Poincar\'{e} invariance of the deformations, combined with the requirement
that the interaction vertices contain at most two spatiotemporal derivatives
of the fields, we investigate the consistent selfinteractions that can be
added to a collection of massless tensor fields with the mixed symmetry $%
(3,1)$ and respectively $(2,2)$. The computations are done with the help of
the deformation theory based on a cohomological approach, in the context of
the antifield-BRST formalism. Our result is that no selfinteractions that
deform the original gauge transformations emerge. In the case of the
collection of $(2,2)$ tensor fields it is possible to add a sum of
cosmological terms to the free Lagrangian.

\end{abstract}

\section{Introduction}

Tensor fields in \textquotedblleft exotic\textquotedblright\ representations
of the Lorentz group, characterized by a mixed Young symmetry type~\cite%
{curt1,curt2,aul,labast1,labast2,burd,zinov1}, held the attention lately on
some important issues, like the dual formulation of field theories of spin
two or higher~\cite%
{dualsp1,dualsp2,dualsp2a,dualsp2b,dualsp3,dualsp4,dualsp5}, the
impossibility of consistent cross-interactions in the dual formulation of
linearized gravity~\cite{lingr}, a Lagrangian first-order approach~\cite%
{zinov2,zinov3} to some classes of massless or partially massive mixed
symmetry type tensor gauge fields, suggestively resembling to the tetrad
formalism of General Relativity, or the derivation of some exotic
gravitational interactions \cite{boulangerCQG,ancoPRD}. An important matter
related to mixed symmetry type tensor fields is the study of their
consistent interactions, among themselves as well as with higher-spin gauge
theories~\cite{high1,high2,high3,high4,7,9,kk,3,4}. The most efficient
approach to this problem is the cohomological one, based on the deformation
of the solution to the master equation~\cite{def}.

The purpose of this paper is to investigate the consistent selfinteractions
in a collection of massless tensor gauge fields with the mixed symmetry of a
two-column Young diagram of the type $(3,1)$, and respectively a collection
of massless tensor gauge fields with the mixed symmetry $(2,2)$. It is worth
mentioning the duality of a free massless tensor gauge field with the mixed
symmetry $\left( 3,1\right) $ to the Pauli--Fierz theory in $D=6$ dimensions
and, in this respect, some developments concerning the dual formulations of
linearized gravity from the perspective of $M$-theory~\cite{mth1,mth2,mth3}.
Our analysis relies on the deformation of the solution to the master
equation by means of cohomological techniques with the help of the local
BRST cohomology, whose component in a single $(3,1)$ sector has been
reported in detail in~\cite{t31jhep}, while in a single $\left( 2,2\right) $
sector has been considered in~\cite{r22,r22th}. Under the hypotheses of
analyticity in the coupling constant, locality, Lorentz covariance, and
Poincar\'{e} invariance of the deformations, combined with the preservation
of the number of derivatives on each field, we find that no selfinteractions
that deform the original gauge transformations emerge. In the case of the
collection of $(2,2)$ tensor fields it is possible to add a sum of
cosmological terms to the free Lagrangian.

\section{Brief review of the deformation procedure}

There are three main types of consistent interactions that can be added to a
given gauge theory: \textit{(i)} the first type deforms only the Lagrangian
action, but not its gauge transformations, \textit{(ii)} the second kind
modifies both the action and its transformations, but not the gauge algebra,
and \textit{(iii)} the third, and certainly most interesting category,
changes everything, namely, the action, its gauge symmetries and the
accompanying algebra.

The reformulation of the problem of consistent deformations of a given
action and of its gauge symmetries in the antifield-BRST setting is based on
the observation that if a deformation of the classical theory can be
consistently constructed, then the solution $S$ to the master equation for
the initial theory can be deformed into the solution $\bar{S}$ to the master
equation for the interacting theory
\begin{eqnarray}
S &\longrightarrow &\bar{S}=S+gS_{1}+g^{2}S_{2}+g^{3}S_{3}+g^{4}S_{4}+\cdots
,  \label{ec39} \\
\left( S,S\right) =0 &\longrightarrow &\left( \bar{S},\bar{S}\right) =0.
\label{ec39b}
\end{eqnarray}%
The projection of (\ref{ec39b}) for $\bar{S}$ on the various powers of the
coupling constant induces the following tower of equations:
\begin{eqnarray}
g^{0} &:&\left( S,S\right) =0,  \label{ec40} \\
g^{1} &:&\left( S_{1},S\right) =0,  \label{ec41} \\
g^{2} &:&\left( S_{2},S\right) +\frac{1}{2}\left( S_{1},S_{1}\right) =0,
\label{ec42} \\
g^{3} &:&\left( S_{3},S\right) +\left( S_{1},S_{2}\right) =0,  \label{ec43}
\\
g^{4} &:&\left( S_{4},S\right) +\left( S_{1},S_{3}\right) +\frac{1}{2}\left(
S_{2},S_{2}\right) =0,  \label{ec44} \\
&&\vdots   \notag
\end{eqnarray}%
The first equation is satisfied by hypothesis. The second one governs the
first-order deformation of the solution to the master equation, $S_{1}$, and
it expresses the fact that $S_{1}$ is a BRST co-cycle, $sS_{1}=0$, and hence
it exists and is local. The remaining equations are responsible for the
higher-order deformations of the solution to the master equation. No
obstructions arise in finding solutions to them as long as no further
restrictions, such as spatiotemporal locality, are imposed. Obviously, only
non-trivial first-order deformations should be considered, since trivial
ones ($S_{1}=sB$) lead to trivial deformations of the initial theory, and
can be eliminated by convenient redefinitions of the fields. Ignoring the
trivial deformations, it follows that $S_{1}$ is a non-trivial
BRST-observable, $S_{1}\in H^{0}\left( s\right) $ (where $H^{0}\left(
s\right) $ denotes the cohomology space of the BRST differential in ghost
number zero). Once the deformation equations ((\ref{ec41})--(\ref{ec44}),
etc.) have been solved by means of specific cohomological techniques, from
the consistent non-trivial deformed solution to the master equation one can
extract all the information on the gauge structure of the resulting
interacting theory.

\section{Selfinteractions for a collection of massless tensor fields with
the mixed symmetry $(3,1)$\label{selfcol(3,1)}}

\subsection{Free model: Lagrangian formulation and BRST symmetry\label{col2}}

The starting point is given by the Lagrangian action for a collection of
free, massless tensor fields with the mixed symmetry $(3,1)$
\begin{eqnarray}
S_{0}^{\mathrm{t}}\left[ t_{\lambda \mu \nu |\alpha }^{A}\right] &=&\int
\left\{ \frac{1}{2}\left[ \left( \partial ^{\rho }t_{A}^{\lambda \mu \nu
|\alpha }\right) \left( \partial _{\rho }t_{\lambda \mu \nu |\alpha
}^{A}\right) -\left( \partial _{\alpha }t_{A}^{\lambda \mu \nu |\alpha
}\right) \left( \partial ^{\beta }t_{\lambda \mu \nu |\beta }^{A}\right) %
\right] \right.  \notag \\
&&-\frac{3}{2}\left[ \left( \partial _{\lambda }t_{A}^{\lambda \mu \nu
|\alpha }\right) \left( \partial ^{\rho }t_{\rho \mu \nu |\alpha
}^{A}\right) +\left( \partial ^{\rho }t_{A}^{\lambda \mu }\right) \left(
\partial _{\rho }t_{\lambda \mu }^{A}\right) \right]  \notag \\
&&\left. +3\left[ \left( \partial _{\alpha }t_{A}^{\lambda \mu \nu |\alpha
}\right) \left( \partial _{\lambda }t_{\mu \nu }^{A}\right) +\left( \partial
_{\rho }t_{A}^{\rho \mu }\right) \left( \partial ^{\lambda }t_{\lambda \mu
}^{A}\right) \right] \right\} d^{D}x,  \label{colt1}
\end{eqnarray}%
in a Minkowski space-time of dimension $D\geq 5$. Everywhere in this paper
we employ the flat Minkowski metric of `mostly plus' signature $\sigma ^{\mu
\nu }=\sigma _{\mu \nu }=(-++++\cdots )$. The uppercase indices $A$, $B$,
etc. stand for the collection indices and are assumed to take discrete
values $1$, $2$, $\ldots $, $N$. They are lowered with a symmetric,
constant, and invertible matrix, of elements $k_{AB}$, and are raised with
the help of the elements $k^{AB}$ of its inverse. Each field $t_{\lambda \mu
\nu |\alpha }^{A}$ is completely antisymmetric in its first three (Lorentz)
indices and satisfies the identity $t_{\left[ \lambda \mu \nu |\alpha \right]
}^{A}\equiv 0$. Here and in the sequel the notation $[\lambda \cdots \alpha
] $ signifies complete antisymmetry with respect to the (Lorentz) indices
between brackets, with the conventions that the minimum number of terms is
always used and the result is never divided by the number of terms. The
notation $t_{\lambda \mu }^{A}$ from (\ref{colt1}) signifies the trace of $%
t_{\lambda \mu \nu |\alpha }^{A}$, defined by $t_{\lambda \mu }^{A}=\sigma
^{\nu \alpha }t_{\lambda \mu \nu |\alpha }^{A}$. The trace components define
an antisymmetric tensor, $t_{\lambda \mu }^{A}=-t_{\mu \lambda }^{A}$. A
generating set of gauge transformations for action (\ref{colt1}) can be
chosen of the form
\begin{eqnarray}
\delta _{\epsilon ,\chi }t_{\lambda \mu \nu |\alpha }^{A} &=&3\partial
_{\alpha }\epsilon _{\lambda \mu \nu }^{A}+\partial _{\left[ \lambda \right.
}\epsilon _{\left. \mu \nu \right] \alpha }^{A}+\partial _{\left[ \lambda
\right. }\chi _{\left. \mu \nu \right] |\alpha }^{A}  \notag \\
&=&-3\partial _{\left[ \lambda \right. }\epsilon _{\left. \mu \nu \alpha %
\right] }^{A}+4\partial _{\left[ \lambda \right. }\epsilon _{\left. \mu \nu %
\right] \alpha }^{A}+\partial _{\left[ \lambda \right. }\chi _{\left. \mu
\nu \right] |\alpha }^{A},  \label{colt7}
\end{eqnarray}%
where the gauge parameters $\epsilon _{\lambda \mu \nu }^{A}$ are completely
antisymmetric, and the gauge parameters $\chi _{\mu \nu |\alpha }^{A}$ (also
bosonic) define a collection of tensor fields with the mixed symmetry $(2,1)$%
. It can be shown \cite{t31jhep} that the generating set (\ref{colt7}) is
off-shell reducible of order two and the associated gauge algebra is
Abelian. Consequently, the Cauchy order of this linear gauge theory is equal
to four.

The most general quantities, invariant under the gauge transformations (\ref%
{colt7}), are given by the components of the curvature tensors associated
with each field from the collection%
\begin{equation}
K_{A}^{\lambda \mu \nu \xi |\alpha \beta }=\partial ^{\alpha }\partial ^{
\left[ \lambda \right. }t_{A}^{\left. \mu \nu \xi \right] |\beta }-\partial
^{\beta }\partial ^{\left[ \lambda \right. }t_{A}^{\left. \mu \nu \xi \right]
|\alpha }  \label{colt20}
\end{equation}%
together with their space-time derivatives. It is easy to check that they
display the mixed symmetry $(4,2)$.

The construction of the BRST symmetry for the free model under study debuts
with the identification of the algebra on which the BRST differential $s$
acts. The ghost spectrum comprises the fermionic ghosts $\left\{ \eta
_{\lambda \mu \nu }^{A},\mathcal{G}_{\mu \nu |\alpha }^{A}\right\} $
respectively associated with the gauge parameters $\left\{ \epsilon
_{\lambda \mu \nu }^{A},\chi _{\mu \nu |\alpha }^{A}\right\} $ from (\ref%
{colt7}), the bosonic ghosts for ghosts $\left\{ C_{\mu \nu }^{A},G_{\nu
\alpha }^{A}\right\} $ due to the first-order reducibility, and the
fermionic ghosts for ghosts for ghosts $C_{\nu }^{A}$ corresponding to the
maximum reducibility order (two). We ask that $\eta _{\lambda \mu \nu }^{A}$
and $C_{\mu \nu }^{A}$ are completely antisymmetric, $\mathcal{G}_{\mu \nu
|\alpha }^{A}$ exhibit the mixed symmetry $(2,1)$, and $G_{\nu \alpha }^{A}$
are symmetric. The antifield spectrum comprises the antifields $t_{A}^{\ast
\lambda \mu \nu |\alpha }$ associated with the original fields and those
corresponding to the ghosts, $\left\{ \eta _{A}^{\ast \lambda \mu \nu },%
\mathcal{G}_{A}^{\ast \mu \nu |\alpha }\right\} $, $\left\{ C_{A}^{\ast \mu
\nu },G_{A}^{\ast \nu \alpha }\right\} $, and $C_{A}^{\ast \nu }$.

Since both the gauge generators and reducibility functions for this model
are field-independent, it follows that the BRST differential $s$ simply
reduces to $s=\delta +\gamma $, where $\delta $ represents the Koszul--Tate
differential, graded by the antighost number $\mathrm{agh}$ ($\mathrm{agh}%
\left( \delta \right) =-1$), and $\gamma $ stands for the exterior
longitudinal differential, whose degree is named pure ghost number $\mathrm{%
pgh}$ ($\mathrm{pgh}\left( \gamma \right) =1$). These two degrees do not
interfere ($\mathrm{agh}\left( \gamma \right) =0$, $\mathrm{pgh}\left(
\delta \right) =0$). The overall degree that grades the BRST complex is
known as the ghost number ($\mathrm{gh}$) and is defined like the difference
between the pure ghost number and the antighost number, such that $\mathrm{gh%
}\left( s\right) =\mathrm{gh}\left( \delta \right) =\mathrm{gh}\left( \gamma
\right) =1$. According to the standard rules of the BRST method, the
corresponding degrees of the generators from the BRST complex are valued like%
\begin{gather*}
\mathrm{pgh}\left( t_{\lambda \mu \nu |\alpha }^{A}\right) =0,\quad \mathrm{%
pgh}\left( \eta _{\lambda \mu \nu }^{A}\right) =\mathrm{pgh}\left( \mathcal{G%
}_{\mu \nu |\alpha }^{A}\right) =1, \\
\mathrm{pgh}\left( C_{\mu \nu }^{A}\right) =\mathrm{pgh}\left( G_{\nu \alpha
}^{A}\right) =2, \\
\mathrm{pgh}\left( t_{A}^{\ast \lambda \mu \nu |\alpha }\right) =\mathrm{pgh}%
\left( \eta _{A}^{\ast \lambda \mu \nu }\right) =\mathrm{pgh}\left( \mathcal{%
G}_{A}^{\ast \mu \nu |\alpha }\right) =\mathrm{pgh}\left( C_{A}^{\ast \mu
\nu }\right) =\mathrm{pgh}\left( G_{A}^{\ast \nu \alpha }\right) =0, \\
\mathrm{agh}\left( t_{\lambda \mu \nu |\alpha }^{A}\right) =\mathrm{agh}%
\left( \eta _{\lambda \mu \nu }^{A}\right) =\mathrm{agh}\left( \mathcal{G}%
_{\mu \nu |\alpha }^{A}\right) =\mathrm{agh}\left( C_{\mu \nu }^{A}\right) =%
\mathrm{agh}\left( G_{\nu \alpha }^{A}\right) =0, \\
\mathrm{agh}\left( t_{A}^{\ast \lambda \mu \nu |\alpha }\right) =1,\quad
\mathrm{agh}\left( \eta _{A}^{\ast \lambda \mu \nu }\right) =\mathrm{agh}%
\left( \mathcal{G}_{A}^{\ast \mu \nu |\alpha }\right) =2, \\
\mathrm{agh}\left( C_{A}^{\ast \mu \nu }\right) =\mathrm{agh}\left(
G_{A}^{\ast \nu \alpha }\right) =3.
\end{gather*}%
The Koszul--Tate differential is imposed to realize a homological resolution
of the algebra of smooth functions defined on the stationary surface of
field equations, while the exterior longitudinal differential is related to
the gauge symmetries (see relations (\ref{colt7})) of action (\ref{colt1})
through its cohomology at pure ghost number zero computed in the cohomology
of $\delta $, which is required to be the algebra of physical observables
for the free model under consideration. The actions of $\delta $ and $\gamma
$ on the generators from the BRST complex, which enforce all the above
mentioned properties, are given by%
\begin{equation}
\gamma t_{\lambda \mu \nu |\alpha }^{A}=-3\partial _{\left[ \lambda \right.
}\eta _{\left. \mu \nu \alpha \right] }^{A}+4\partial _{\left[ \lambda
\right. }\eta _{\left. \mu \nu \right] \alpha }^{A}+\partial _{\left[
\lambda \right. }\mathcal{G}_{\left. \mu \nu \right] |\alpha }^{A},
\label{colt49}
\end{equation}%
\begin{equation}
\gamma \eta _{\lambda \mu \nu }^{A}=-\frac{1}{2}\partial _{\left[ \lambda
\right. }C_{\left. \mu \nu \right] }^{A},  \label{colt50}
\end{equation}%
\begin{equation}
\gamma \mathcal{G}_{\mu \nu |\alpha }^{A}=2\partial _{\left[ \mu \right.
}C_{\left. \nu \alpha \right] }^{A}-3\partial _{\left[ \mu \right.
}C_{\left. \nu \right] \alpha }^{A}+\partial _{\left[ \mu \right. }G_{\left.
\nu \right] \alpha }^{A},  \label{colt51}
\end{equation}%
\begin{equation}
\gamma C_{\mu \nu }^{A}=\partial _{\left[ \mu \right. }C_{\left. \nu \right]
}^{A},\qquad \gamma G_{\nu \alpha }^{A}=-3\partial _{\left( \nu \right.
}C_{\left. \alpha \right) }^{A},\qquad \gamma C_{\nu }^{A}=0,  \label{colt52}
\end{equation}%
\begin{equation}
\gamma t_{A}^{\ast \lambda \mu \nu |\alpha }=\gamma \eta _{A}^{\ast \lambda
\mu \nu }=\gamma \mathcal{G}_{A}^{\ast \mu \nu |\alpha }=\gamma C_{A}^{\ast
\mu \nu }=\gamma G_{A}^{\ast \nu \alpha }=\gamma C_{A}^{\ast \nu }=0,
\label{colt53}
\end{equation}%
\begin{equation}
\delta t_{\lambda \mu \nu |\alpha }^{A}=\delta \eta _{\lambda \mu \nu
}^{A}=\delta \mathcal{G}_{\mu \nu |\alpha }^{A}=\delta C_{\mu \nu
}^{A}=\delta G_{\nu \alpha }^{A}=\delta C_{\nu }^{A}=0,  \label{colt54}
\end{equation}%
\begin{equation}
\delta t_{A}^{\ast \lambda \mu \nu |\alpha }=T_{A}^{\lambda \mu \nu |\alpha
},\qquad \delta \eta _{A}^{\ast \lambda \mu \nu }=-4\partial _{\alpha
}t_{A}^{\ast \lambda \mu \nu |\alpha },  \label{colt55}
\end{equation}%
\begin{equation}
\delta \mathcal{G}_{A}^{\ast \mu \nu |\alpha }=-\partial _{\lambda }\left(
3t_{A}^{\ast \lambda \mu \nu |\alpha }-t_{A}^{\ast \mu \nu \alpha |\lambda
}\right) ,  \label{colt56}
\end{equation}%
\begin{equation}
\delta C_{A}^{\ast \mu \nu }=3\partial _{\lambda }\left( \mathcal{G}%
_{A}^{\ast \mu \nu |\lambda }-\frac{1}{2}\eta _{A}^{\ast \lambda \mu \nu
}\right) ,\qquad \delta G_{A}^{\ast \nu \alpha }=\partial _{\mu }\mathcal{G}%
_{A}^{\ast \mu \left( \nu |\alpha \right) },  \label{colt57}
\end{equation}%
\begin{equation}
\delta C_{A}^{\ast \nu }=6\partial _{\mu }\left( G_{A}^{\ast \mu \nu }-\frac{%
1}{3}C_{A}^{\ast \mu \nu }\right) ,  \label{colt58}
\end{equation}%
where $T_{A}^{\lambda \mu \nu |\alpha }=-\delta S_{0}^{\mathrm{t}}/\delta
t_{\lambda \mu \nu |\alpha }^{A}$ reads%
\begin{eqnarray}
T_{A}^{\lambda \mu \nu |\alpha } &=&\Box t_{A}^{\lambda \mu \nu |\alpha
}-\partial _{\rho }\left( \partial ^{\left[ \lambda \right. }t_{A}^{\left.
\mu \nu \right] \rho |\alpha }+\partial ^{\alpha }t_{A}^{\lambda \mu \nu
|\rho }\right) +\partial ^{\alpha }\partial ^{\left[ \lambda \right.
}t_{A}^{\left. \mu \nu \right] }  \notag \\
&&+\sigma ^{\alpha \left[ \lambda \right. }\left( \partial _{\rho }\left(
\partial _{\beta }t_{A}^{\left. \mu \nu \right] \rho |\beta }-\partial ^{\mu
}t_{A}^{\left. \nu \right] \rho }\right) -\Box t_{A}^{\left. \mu \nu \right]
}\right) .  \label{colt17}
\end{eqnarray}%
By convention, we take $\delta $ and $\gamma $ to act like right
derivations. We note that the action of the Koszul--Tate differential on the
antifields with the antighost number equal to two and respectively three
gains a simpler expression if we perform the changes of variables%
\begin{equation}
\mathcal{G}_{A}^{\prime \ast \mu \nu |\alpha }=\mathcal{G}_{A}^{\ast \mu \nu
|\alpha }+\frac{1}{4}\eta _{A}^{\ast \mu \nu \alpha },\qquad G_{A}^{\prime
\ast \nu \alpha }=G_{A}^{\ast \nu \alpha }-\frac{1}{3}C_{A}^{\ast \nu \alpha
}.  \label{colt58a}
\end{equation}%
The antifields $\mathcal{G}_{A}^{\prime \ast \mu \nu |\alpha }$ are still
antisymmetric in their first two indices, but do not fulfill the identity $%
\mathcal{G}_{A}^{\prime \ast \left[ \mu \nu |\alpha \right] }\equiv 0$, and $%
G_{A}^{\prime \ast \nu \alpha }$ have no definite symmetry or antisymmetry
properties. With the help of relations (\ref{colt55})--(\ref{colt58}), we
find that $\delta $ acts on the transformed antifields through the relations%
\begin{equation}
\delta \mathcal{G}_{A}^{\prime \ast \mu \nu |\alpha }=-3\partial _{\lambda
}t_{A}^{\ast \lambda \mu \nu |\alpha },\qquad \delta G_{A}^{\prime \ast \nu
\alpha }=2\partial _{\mu }\mathcal{G}_{A}^{\prime \ast \mu \nu |\alpha
},\qquad \delta C_{A}^{\ast \nu }=6\partial _{\mu }G_{A}^{\prime \ast \mu
\nu }.  \label{colt58b}
\end{equation}%
The same observation is valid with respect to $\gamma $ if we make the
changes
\begin{equation}
\mathcal{G}_{\mu \nu |\alpha }^{\prime A}=\mathcal{G}_{\mu \nu |\alpha
}^{A}+4\eta _{\mu \nu \alpha }^{A},\qquad G_{\nu \alpha }^{\prime A}=G_{\nu
\alpha }^{A}-3C_{\nu \alpha }^{A},  \label{colt58ba}
\end{equation}%
in terms of which we can write
\begin{equation}
\gamma t_{\lambda \mu \nu |\alpha }^{A}=-\frac{1}{4}\partial _{\left[
\lambda \right. }\mathcal{G}_{\left. \mu \nu |\alpha \right] }^{\prime
A}+\partial _{\left[ \lambda \right. }\mathcal{G}_{\left. \mu \nu \right]
|\alpha }^{\prime A},\ \gamma \mathcal{G}_{\mu \nu |\alpha }^{\prime
A}=\partial _{\left[ \mu \right. }G_{\left. \nu \right] \alpha }^{\prime
A},\ \gamma G_{\nu \alpha }^{\prime A}=-6\partial _{\nu }C_{\alpha }^{A}.
\label{colt58bd}
\end{equation}%
Again, $\mathcal{G}_{\mu \nu |\alpha }^{\prime A}$ are antisymmetric in
their first two indices, but do not satisfy the identity $\mathcal{G}_{\left[
\mu \nu |\alpha \right] }^{\prime A}\equiv 0$, while $G_{\nu \alpha
}^{\prime A}$ have no definite symmetry or antisymmetry. We have
deliberately chosen the same notations for the transformed variables (\ref%
{colt58a}) and (\ref{colt58ba}) since they actually form pairs that are
conjugated in the antibracket%
\begin{eqnarray*}
\left( \mathcal{G}_{\mu \nu |\alpha }^{\prime A},\mathcal{G}_{B}^{\prime
\ast \mu _{1}\nu _{1}|\alpha _{1}}\right)  &=&\frac{1}{2}\delta
_{B}^{A}\delta _{\mu }^{\left[ \mu _{1}\right. }\delta _{\nu }^{\left. \nu
_{1}\right] }\delta _{\alpha }^{\alpha _{1}}, \\
\left( G_{\nu \alpha }^{\prime A},G_{B}^{\prime \ast \nu _{1}\alpha
_{1}}\right)  &=&\delta _{B}^{A}\delta _{\nu }^{\nu _{1}}\delta _{\alpha
}^{\alpha _{1}}.
\end{eqnarray*}

The Lagrangian BRST differential admits a canonical action in a structure
named antibracket and defined by decreeing the fields/ghosts conjugated with
the corresponding antifields, $s\cdot =\left( \cdot ,S\right) $, where $%
\left( ,\right) $ signifies the antibracket and $S$ denotes the canonical
generator of the BRST symmetry. It is a bosonic functional of ghost number
zero, involving both field/ghost and antifield spectra, that obeys the
master equation $\left( S,S\right) =0$. The master equation is equivalent
with the second-order nilpotency of $s$, where its solution $S$ encodes the
entire gauge structure of the associated theory. Taking into account
formulas (\ref{colt49})--(\ref{colt58}) as well as the standard actions of $%
\delta $ and $\gamma $ in canonical form, we find that the complete solution
to the master equation for the free model under study is given by%
\begin{eqnarray}
S^{\mathrm{t}} &=&S_{0}^{\mathrm{t}}\left[ t_{\lambda \mu \nu |\alpha }^{A}%
\right] +\int \left( t_{A}^{\ast \lambda \mu \nu |\alpha }\left( 3\partial
_{\alpha }\eta _{\lambda \mu \nu }^{A}+\partial _{\left[ \lambda \right.
}\eta _{\left. \mu \nu \right] \alpha }^{A}+\partial _{\left[ \lambda
\right. }\mathcal{G}_{\left. \mu \nu \right] |\alpha }^{A}\right) \right.
\notag \\
&&-\frac{1}{2}\eta _{A}^{\ast \lambda \mu \nu }\partial _{\left[ \lambda
\right. }C_{\left. \mu \nu \right] }^{A}+\mathcal{G}_{A}^{\ast \mu \nu
|\alpha }\left( 2\partial _{\alpha }C_{\mu \nu }^{A}-\partial _{\left[ \mu
\right. }C_{\left. \nu \right] \alpha }^{A}+\partial _{\left[ \mu \right.
}G_{\left. \nu \right] \alpha }^{A}\right)  \notag \\
&&\left. +C_{A}^{\ast \mu \nu }\partial _{\left[ \mu \right. }C_{\left. \nu %
\right] }^{A}-3G_{A}^{\ast \nu \alpha }\partial _{\left( \nu \right.
}C_{\left. \alpha \right) }^{A}\right) d^{D}x.  \label{colt60}
\end{eqnarray}

\subsection{Computation of basic cohomologies\label{col5}}

In order to analyze equation (\ref{ec41}) (that governs the first-order
deformation) we make the notation $S_{1}=\int a^{\mathrm{t}}d^{D}x$ and
write this equation in its local form and in dual notations, $sa^{\mathrm{t}%
}=\partial _{\mu }m_{\mathrm{t}}^{\mu }$. Now, we approach the last equation
in a standard manner, namely, we develop $a^{\mathrm{t}}$ according to the
antighost number and assume that this expansion contains a finite number of
terms, of maximum antighost number $I$. In order to ensure the space-time
locality of the deformations, from now on we work in the algebra of local
differential forms with coefficients that are polynomial functions in the
fields, ghosts, antifields, and their space-time derivatives (algebra of
local forms). This means that we assume the non-integrated density of the
first-order deformation, $a^{\mathrm{t}}$, to be a polynomial function in
all these variables (algebra of local functions).

By taking into account the splitting $s=\delta +\gamma $ of the BRST
differential, the equation $sa^{\mathrm{t}}=\partial _{\mu }m_{\mathrm{t}%
}^{\mu }$ becomes equivalent to a tower of local equations, corresponding to
the different decreasing values of the antighost number%
\begin{eqnarray}
\gamma a_{I}^{\mathrm{t}} &=&\partial _{\mu }\overset{(I)}{m}_{\mathrm{t}%
}^{\mu },  \label{colt65f} \\
\delta a_{I}^{\mathrm{t}}+\gamma a_{I-1}^{\mathrm{t}} &=&\partial _{\mu }%
\overset{(I-1)}{m}_{\mathrm{t}}^{\mu },  \label{colt65d} \\
\delta a_{k}^{\mathrm{t}}+\gamma a_{k-1}^{\mathrm{t}} &=&\partial _{\mu }%
\overset{(k-1)}{m}_{\mathrm{t}}^{\mu },\qquad I-1\geq k\geq 1,
\label{colt65e}
\end{eqnarray}%
where $\left( \overset{(k)}{m}_{\mathrm{t}}^{\mu }\right) _{k=\overline{0,I}%
} $ are some local currents, with $\mathrm{agh}\left( \overset{(k)}{m}_{%
\mathrm{t}}^{\mu }\right) =k$. It can be proved that we can replace equation
(\ref{colt65f}) at strictly positive antighost numbers with the homogeneous
equation%
\begin{equation}
\gamma a_{I}^{\mathrm{t}}=0,\qquad I>0.  \label{coltomI}
\end{equation}%
The proof can be done like in the Appendix A, Corollary 1, from \cite%
{t31jhep}. In conclusion, under the assumption that $I>0$, the
representative of highest antighost number from the non-integrated density
of the first-order deformation can always be taken to be $\gamma $-closed,
such that equation $sa^{\mathrm{t}}=\partial _{\mu }m_{\mathrm{t}}^{\mu }$,
associated with the local form of the first-order deformation equation, is
completely equivalent to the tower of equations given by (\ref{coltomI}) and
(\ref{colt65d})--(\ref{colt65e}).

Before proceeding to the analysis of the solutions to the first-order
deformation equation, let us briefly comment on the uniqueness and
triviality of such solutions. Due to the second-order nilpotency of $\gamma $
($\gamma ^{2}=0$), the solution to the top equation, (\ref{coltomI}), is
clearly unique up to $\gamma $-exact contributions, $a_{I}^{\mathrm{t}%
}\rightarrow a_{I}^{\mathrm{t}}+\gamma b_{I}$. Meanwhile, if $a_{I}^{\mathrm{%
t}}$ reduces to $\gamma $-exact terms only, $a_{I}^{\mathrm{t}}=\gamma b_{I}$%
, then it can be made to vanish, $a_{I}^{\mathrm{t}}=0$. In other words, the
non-triviality of the first-order deformation $a^{\mathrm{t}}$ is translated
at its highest antighost number component into the requirement that $a_{I}^{%
\mathrm{t}}\in H^{I}\left( \gamma \right) $, where $H^{I}\left( \gamma
\right) $ denotes the cohomology of the exterior longitudinal differential $%
\gamma $ in pure ghost number equal to $I$ computed in the algebra of local
functions. At the same time, the general condition on the non-integrated
density of the first-order deformation to generate an element $a^{\mathrm{t}%
}d^{D}x$ from a non-trivial cohomological class of $H^{0,D}\left( s|d\right)
$ (the local cohomology of the BRST differential $s$ --- where $d$ means the
exterior space-time differential --- in ghost number zero and in maximum
form degree, computed in the algebra of local forms) shows on the one hand
that the solution to equation $sa^{\mathrm{t}}=\partial _{\mu }m_{\mathrm{t}%
}^{\mu }$ is unique up to $s$-exact pieces plus total derivatives and, on
the other hand, that if the general solution to this equation is completely
trivial, $a^{\mathrm{t}}=sb+\partial _{\mu }n^{\mu }$, then it can be made
to vanish, $a^{\mathrm{t}}=0$.

We have seen that the solution to equation (\ref{coltomI}) belongs to the
cohomology of the exterior longitudinal differential computed in the algebra
of local functions, such that we need to compute $H^{\ast }\left( \gamma
\right) $ in order to construct the component of highest antighost number
from the first-order deformation. We will see that we also need to compute
the characteristic cohomology $H_{I}^{D}\left( \delta |d\right) $ (the local
cohomology of the Koszul--Tate differential $\delta $ in antighost number $I$
and in maximum form degree, computed in the algebra of local forms with the
pure ghost number equal to zero).

Acting like in \cite{t31jhep}, it is easy to see that $H^{\ast }\left(
\gamma \right) $ is generated by the quantities
\begin{equation}
\begin{array}{ccc}
\mathrm{pgh} & \mathrm{BRST\;generator} & \mathrm{non-trivial\
objects\;from\;}H^{\ast }\left( \gamma \right) \\
0 & \left\{
\begin{array}{c}
\Pi ^{\ast \Delta },\partial \Pi ^{\ast \Delta },\ldots \\
t_{\lambda \mu \nu |\alpha }^{A},\partial t_{\lambda \mu \nu |\alpha
}^{A},\ldots%
\end{array}%
\right. & \left\{
\begin{array}{c}
\Pi ^{\ast \Delta },\partial \Pi ^{\ast \Delta },\ldots \\
K_{\lambda \mu \nu \xi |\alpha \beta }^{A},\partial K_{\lambda \mu \nu \xi
|\alpha \beta }^{A},\ldots%
\end{array}%
\right. \\
1 & \left\{
\begin{array}{l}
\eta _{\lambda \mu \nu }^{A},\partial \eta _{\lambda \mu \nu }^{A},\ldots \\
\mathcal{G}_{\mu \nu |\alpha }^{A},\partial \mathcal{G}_{\mu \nu |\alpha
}^{A},\ldots%
\end{array}%
\right. &
\begin{array}{l}
\mathcal{F}_{\lambda \mu \nu \alpha }^{A}=\partial _{\left[ \lambda \right.
}\eta _{\left. \mu \nu \alpha \right] }^{A},%
\end{array}
\\
2 & \left\{
\begin{array}{l}
C_{\mu \nu }^{A},\partial C_{\mu \nu }^{A},\ldots \\
G_{\nu \alpha }^{A},\partial G_{\nu \alpha }^{A},\ldots%
\end{array}%
\right. & \mathrm{-} \\
3 & C_{\nu }^{A},\partial C_{\nu }^{A},\ldots & C_{\nu }^{A}%
\end{array}%
,  \label{coltablegama}
\end{equation}%
where $\Pi ^{\ast \Delta }$ is a generic notation for all the antifields.
So, the most general, non-trivial solution to the equation (\ref{coltomI})
(up to trivial, $\gamma $-exact contributions) reads
\begin{equation}
a_{I}^{\mathrm{t}}=\alpha _{I}\left( \left[ \Pi ^{\ast \Delta }\right] ,%
\left[ K_{\lambda \mu \nu \xi |\alpha \beta }^{A}\right] \right) \omega
^{I}\left( \mathcal{F}_{\lambda \mu \nu \alpha }^{A},C_{\nu }^{A}\right) .
\label{colt81}
\end{equation}%
The notation $f([q])$ means that $f$ depends on $q$ and its derivatives up
to a finite order, while $\omega ^{I}$ denotes the elements of pure ghost
number $I$ (and antighost number zero) of a basis in the space of
polynomials in $\mathcal{F}_{\lambda \mu \nu \alpha }^{A}$ and $C_{\nu }^{A}$%
, which is finite dimensional since these variables anticommute. The objects
$\alpha _{I}$ (obviously non-trivial in $H^{0}\left( \gamma \right) $) were
taken to have a bounded number of derivatives, and therefore they are
polynomials in the antifields $\Theta ^{\ast \Delta }$, in the curvature
tensors $K_{\lambda \mu \nu \xi |\alpha \beta }^{A}$, as well as in their
derivatives. They are nothing but the invariant polynomials of the theory
described by formulas (\ref{colt1})--(\ref{colt7}) in form degree equal to
zero. At zero antighost number, the invariant polynomials are polynomials in
the curvature tensors $K_{\lambda \mu \nu \xi |\alpha \beta }^{A}$ and in
their derivatives.

Replacing solution (\ref{colt81}) into equation (\ref{colt65d}) and taking
into account definitions (\ref{colt54})--(\ref{colt58}), we remark that a
necessary (but not sufficient) condition for the existence of (non-trivial)
solutions $a_{I-1}^{\mathrm{t}}$ is that the invariant polynomials $\alpha
_{I}$ generate (non-trivial) objects from the characteristic cohomology $%
H_{I}^{D}\left( \delta |d\right) $ in antighost number $I>0$, maximum form
degree, and pure ghost number equal to zero\footnote{\label{local}We recall
that the local cohomology $H_{\ast }^{D}\left( \delta |d\right) $ is
completely trivial at both strictly positive antighost \textit{and} pure
ghost numbers (for instance, see~\cite{gen1}, Theorem 5.4 and~\cite{commun1}%
).}, $\alpha _{I}d^{D}x\in H_{I}^{D}\left( \delta |d\right) $. As the free
model under study is a linear gauge theory of Cauchy order equal to four,
the general results from~\cite{gen1} ensure that the entire characteristic
cohomology is trivial in antighost numbers strictly greater than its Cauchy
order%
\begin{equation}
H_{I}^{D}\left( \delta |d\right) =0,\qquad I>4.  \label{colt83}
\end{equation}%
Moreover, it is possible to show that the above result remains valid also in
the algebra of invariant polynomials
\begin{equation}
H_{I}^{\mathrm{inv}D}\left( \delta |d\right) =0,\qquad I>4,  \label{colt83b}
\end{equation}%
where $H_{I}^{\mathrm{inv}D}\left( \delta |d\right) $ is known as the
invariant characteristic cohomology. Looking at the definitions (\ref%
{colt58b}) involving the transformed antifields (\ref{colt58a}), we can
organize the non-trivial, Poincar\'{e}-invariant representatives of $%
H_{I}^{D}\left( \delta |d\right) $ and $H_{I}^{\mathrm{inv}D}\left( \delta
|d\right) $ (for $I\geq 2$) like:
\begin{equation}
\begin{array}{cc}
\mathrm{agh} & H_{I}^{D}\left( \delta |d\right) \ \mathrm{and}\ H_{I}^{%
\mathrm{inv}D}\left( \delta |d\right) \\
I>4 & \mathrm{-} \\
I=4 & f_{\nu }^{A}C_{A}^{\ast \nu }d^{D}x \\
I=3 & f_{\nu \alpha }^{A}G_{A}^{\prime \ast \nu \alpha }d^{D}x \\
I=2 & f_{\mu \nu \alpha }^{A}\mathcal{G}_{A}^{\prime \ast \mu \nu |\alpha
}d^{D}x%
\end{array}%
,  \label{coltabledelta}
\end{equation}%
where all the coefficients denoted by $f$ define some constant,
non-derivative tensors. We remark that in $\left( H_{I}^{D}\left( \delta
|d\right) \right) _{I\geq 2}$ and $\left( H_{I}^{\mathrm{inv}D}\left( \delta
|d\right) \right) _{I\geq 2}$ there is no non-trivial element that
effectively involves the curvatures $K_{\lambda \mu \nu \xi |\alpha \beta
}^{A}$ and/or their derivatives, and the same stands for the quantities that
are more than linear in the antifields and/or depend on their derivatives.
In principle, one can construct from the above elements in (\ref%
{coltabledelta}) other non-trivial invariant polynomials from $%
H_{I}^{D}\left( \delta |d\right) $ or $H_{I}^{\mathrm{inv}D}\left( \delta
|d\right) $, that depend on the space-time co-ordinates. For instance, it
can be checked by direct computation that $\mathcal{G}_{A}^{\prime \ast \mu
\nu |\alpha }f_{\mu \nu \alpha \rho }^{A}x^{\rho }d^{D}x$, with $f_{\mu \nu
\alpha \rho }^{A}$ some completely antisymmetric and constant tensors,
generate non-trivial representatives from both $H_{2}^{D}\left( \delta
|d\right) $ and $H_{2}^{\mathrm{inv}D}\left( \delta |d\right) $. However, we
will discard such candidates as they would break the Poincar\'{e} invariance
of the deformations. In contrast to the groups $\left( H_{I}^{D}\left(
\delta |d\right) \right) _{I\geq 2}$ and $\left( H_{I}^{\mathrm{inv}D}\left(
\delta |d\right) \right) _{I\geq 2}$, which are finite-dimensional, the
cohomology $H_{1}^{D}\left( \delta |d\right) $ at pure ghost number zero,
that is related to global symmetries and ordinary conservation laws, is
infinite-dimensional since the theory is free.

The previous results on $H_{I}^{D}\left( \delta |d\right) $ and $H_{I}^{%
\mathrm{inv}D}\left( \delta |d\right) $ are important because they control
the obstructions to removing the antifields from the first-order
deformation. Indeed, due to (\ref{colt83b}), it follows that we can
successively eliminate all the pieces with $I>4$ from the non-integrated
density of the first-order deformation by adding only trivial terms (the
proof is similar to that from the Appendix C in \cite{t31jhep}), so we can
take, without loss of non-trivial objects, the condition $I\leq 4$ in the
first-order deformation. The last representative is of the form (\ref{colt81}%
), where the invariant polynomials necessarily generate non-trivial objects
from $H_{I}^{\mathrm{inv}D}\left( \delta |d\right) $ if $I=2,3,4$ and
respectively from $H_{1}^{D}\left( \delta |d\right) $ if $I=1$.

\subsection{Cohomological analysis of selfinteractions\label{col5.2}}

Assuming $I=4$, the non-integrated density of the first-order deformation
becomes
\begin{equation}
a^{\mathrm{t}}=a_{0}^{\mathrm{t}}+a_{1}^{\mathrm{t}}+a_{2}^{\mathrm{t}%
}+a_{3}^{\mathrm{t}}+a_{4}^{\mathrm{t}},  \label{colt86}
\end{equation}%
with $a_{4}^{\mathrm{t}}$ ($\gamma a_{4}^{\mathrm{t}}=0$) a non-trivial
element from $H^{4}\left( \gamma \right) $, and hence of the form (see (\ref%
{colt81}))
\begin{equation}
a_{4}^{\mathrm{t}}=\alpha _{4}\omega ^{4}\left( \mathcal{F}_{\lambda \mu \nu
\alpha }^{A},C_{\nu }^{A}\right) ,  \label{colt87}
\end{equation}%
and $\alpha _{4}d^{D}x$ a non-trivial object from $H_{4}^{\mathrm{inv}%
D}\left( \delta |d\right) $. Since the elements of pure ghost number equal
to four from the basis in the space of polynomials in $\mathcal{F}_{\lambda
\mu \nu \alpha }^{A}$ and $C_{\nu }^{A}$ are spanned by the combinations
\begin{equation}
\omega ^{4}:\left( \mathcal{F}_{\lambda \mu \nu \alpha }^{B}C_{\beta }^{C},%
\mathcal{F}_{\lambda \mu \nu \alpha }^{B}\mathcal{F}_{\lambda _{1}\mu
_{1}\nu _{1}\alpha _{1}}^{C}\mathcal{F}_{\lambda _{2}\mu _{2}\nu _{2}\alpha
_{2}}^{D}\mathcal{F}_{\lambda _{3}\mu _{3}\nu _{3}\alpha _{3}}^{E}\right) ,
\label{colt88}
\end{equation}%
with $\mathcal{F}_{\lambda \mu \nu \alpha }^{A}$ given in (\ref{coltablegama}%
), and the non-trivial representatives of the space $H_{4}^{\mathrm{inv}%
D}\left( \delta |d\right) $ are generated by the antifields $C_{\rho }^{\ast
A}$ (see (\ref{coltabledelta})), we obtain that the general form of the last
term from the first-order deformation in the case $I=4$ reads
\begin{eqnarray}
a_{4}^{\mathrm{t}} &=&C_{\rho }^{\ast A}\left( f_{1ABCDE}^{\rho \lambda \mu
\nu \alpha \lambda _{1}\mu _{1}\nu _{1}\alpha _{1}\lambda _{2}\mu _{2}\nu
_{2}\alpha _{2}\lambda _{3}\mu _{3}\nu _{3}\alpha _{3}}\mathcal{F}_{\lambda
\mu \nu \alpha }^{B}\mathcal{F}_{\lambda _{1}\mu _{1}\nu _{1}\alpha _{1}}^{C}%
\mathcal{F}_{\lambda _{2}\mu _{2}\nu _{2}\alpha _{2}}^{D}\mathcal{F}%
_{\lambda _{3}\mu _{3}\nu _{3}\alpha _{3}}^{E}\right.  \notag \\
&&\left. +f_{2ABC}^{\rho \lambda \mu \nu \alpha \beta }\mathcal{F}_{\lambda
\mu \nu \alpha }^{B}C_{\beta }^{C}\right) ,  \label{colt89}
\end{eqnarray}%
where the coefficients denoted by $f$ are some non-derivative constant
tensors. The first term from the right-hand side of (\ref{colt89}) (those
containing homogeneous polynomials of degree four in the ghosts $\mathcal{F}%
_{\lambda \mu \nu \alpha }^{A}$), even if consistent, would lead to
interaction vertices (in the corresponding $a_{0}^{\mathrm{t}}$) of order
five in the space-time derivatives of the fields, which disagrees with the
hypothesis on the maximum derivative order of the interacting Lagrangian to
be equal to two. For this reason, we eliminate this term from $a_{4}^{%
\mathrm{t}}$ by setting the associated coefficient to be equal to zero%
\begin{equation}
f_{1ABCDE}^{\rho \lambda \mu \nu \alpha \lambda _{1}\mu _{1}\nu _{1}\alpha
_{1}\lambda _{2}\mu _{2}\nu _{2}\alpha _{2}\lambda _{3}\mu _{3}\nu
_{3}\alpha _{3}}=0,  \label{colt89a}
\end{equation}%
such that
\begin{equation}
a_{4}^{\mathrm{t}}=f_{2ABC}^{\rho \lambda \mu \nu \alpha \beta }C_{\rho
}^{\ast A}\mathcal{F}_{\lambda \mu \nu \alpha }^{B}C_{\beta }^{C}.
\label{colt90}
\end{equation}%
The requirements that the deformations are manifestly covariant and Poincar%
\'{e} invariant, the fact that we work in space-time dimensions $D\geq 5$,
and the complete antisymmetry of $\mathcal{F}_{\lambda \mu \nu \alpha }^{B}$%
, provide a single non-trivial candidate, namely%
\begin{equation}
D=6,\qquad f_{2ABC}^{\rho \lambda \mu \nu \alpha \beta }=c_{ABC}\varepsilon
^{\rho \lambda \mu \nu \alpha \beta },  \label{colt92}
\end{equation}%
with $c_{ABC}$ some real, arbitrary constants and $\varepsilon ^{\rho
\lambda \mu \nu \alpha \beta }$ the six-dimensional Levi--Civita symbol. As
a consequence, we obtain%
\begin{equation}
a_{4}^{\mathrm{t}}=c_{ABC}\varepsilon ^{\rho \lambda \mu \nu \alpha \beta
}C_{\rho }^{\ast A}\mathcal{F}_{\lambda \mu \nu \alpha }^{B}C_{\beta }^{C}.
\label{colt91}
\end{equation}%
If (\ref{colt91}) is consistent, then it will produce a Lagrangian density
at order one in the coupling constant, $a_{0}^{\mathrm{t}}$, which breaks
the PT invariance.

We will show that solution (\ref{colt91}) is not consistent in antighost
number two, meaning that it cannot provide a solution $a_{2}^{\mathrm{t}}$
to the equation (\ref{colt65e}) for $k=3$. In view of this, we compute the
remaining components from (\ref{colt86}), which are subject to equations (%
\ref{colt65d})--(\ref{colt65e}) for $I=4$%
\begin{eqnarray}
\delta a_{4}^{\mathrm{t}}+\gamma a_{3}^{\mathrm{t}} &=&\partial _{\mu }%
\overset{(3)}{m}_{\mathrm{t}}^{\mu },\qquad \delta a_{3}^{\mathrm{t}}+\gamma
a_{2}^{\mathrm{t}}=\partial _{\mu }\overset{(2)}{m}_{\mathrm{t}}^{\mu },
\label{colt92a} \\
\delta a_{2}^{\mathrm{t}}+\gamma a_{1}^{\mathrm{t}} &=&\partial _{\mu }%
\overset{(1)}{m}_{\mathrm{t}}^{\mu },\qquad \delta a_{1}^{\mathrm{t}}+\gamma
a_{0}^{\mathrm{t}}=\partial _{\mu }\overset{(0)}{m}_{\mathrm{t}}^{\mu }.
\label{colt92b}
\end{eqnarray}%
Replacing (\ref{colt91}) into the former equation from (\ref{colt92a}) and
using the first definition from (\ref{colt58b}), together with the results
\begin{eqnarray}
\partial _{\rho }\mathcal{F}_{\lambda \mu \nu \alpha }^{A} &=&\gamma \left(
\frac{1}{3}\partial _{\left[ \lambda \right. }t_{\left. \mu \nu \alpha %
\right] |\rho }^{A}\right) ,  \label{colt74} \\
\partial _{\mu }C_{\nu }^{A} &=&\gamma \left( -\frac{1}{6}G_{\mu \nu
}^{\prime A}\right) ,  \label{colt79}
\end{eqnarray}%
we find that%
\begin{equation}
a_{3}^{\mathrm{t}}=-c_{ABC}\varepsilon ^{\rho \lambda \mu \nu \alpha \beta
}G_{\;\;\;\;\;\;\;\rho }^{\prime \ast A\gamma }\left( 8\left( \partial
_{\lambda }t_{\mu \nu \alpha |\gamma }^{B}\right) C_{\beta }^{C}+\mathcal{F}%
_{\lambda \mu \nu \alpha }^{B}G_{\gamma \beta }^{\prime C}\right) ,
\label{colt93}
\end{equation}%
where $G_{\gamma \beta }^{\prime C}$ reads as in (\ref{colt58ba}). In order
to solve the latter equation from (\ref{colt92a}), we initially compute $%
\delta a_{3}^{\mathrm{t}}$ starting with (\ref{colt93}) and using the second
definition from (\ref{colt58b}), and then manipulate the resulting
expression based on formulas (\ref{colt74}), (\ref{colt79}), and the second
relation from (\ref{colt58bd}), obtaining in the end%
\begin{eqnarray}
\delta a_{3}^{\mathrm{t}} &=&\partial _{\sigma }\left( 2c_{ABC}\varepsilon
^{\rho \lambda \mu \nu \alpha \beta }\mathcal{G}_{\;\;\;\;\;\;\;\;\;\;\rho
}^{\prime \ast A\sigma \gamma |}\left( 8\left( \partial _{\lambda }t_{\mu
\nu \alpha |\gamma }^{B}\right) C_{\beta }^{C}+\mathcal{F}_{\lambda \mu \nu
\alpha }^{B}G_{\gamma \beta }^{\prime C}\right) \right)  \notag \\
&&+\gamma \left( -c_{ABC}\varepsilon ^{\rho \lambda \mu \nu \alpha \beta }%
\mathcal{G}_{\;\;\;\;\;\;\;\;\;\;\rho }^{\prime \ast A\sigma \gamma |}\left(
\frac{4}{3}\left( \partial _{\lambda }t_{\mu \nu \alpha |\left[ \sigma
\right. }^{B}\right) G_{\left. \gamma \right] \beta }^{\prime C}+\mathcal{F}%
_{\lambda \mu \nu \alpha }^{B}\mathcal{G}_{\sigma \gamma |\beta }^{\prime
C}\right) \right)  \notag \\
&&-2c_{ABC}\varepsilon ^{\rho \lambda \mu \nu \alpha \beta }\mathcal{G}%
_{\;\;\;\;\;\;\;\;\;\;\rho }^{\prime \ast A\sigma \gamma |}K_{\lambda \mu
\nu \alpha |\sigma \gamma }^{B}C_{\beta }^{C},  \label{colt94}
\end{eqnarray}%
where $K_{\lambda \mu \nu \alpha |\sigma \gamma }^{B}$ is precisely the
curvature tensor (see (\ref{colt20})) and the transformed ghosts $\mathcal{G}%
_{\sigma \gamma |\beta }^{\prime C}$ are defined in (\ref{colt58ba}).
Comparing the latter equation from (\ref{colt92a}) with (\ref{colt94}), we
observe that $a_{3}^{\mathrm{t}}$ of the form (\ref{colt93}) provides a
consistent $a_{2}^{\mathrm{t}}$ if and only if
\begin{equation}
-2c_{ABC}\varepsilon ^{\rho \lambda \mu \nu \alpha \beta }\mathcal{G}%
_{\;\;\;\;\;\;\;\;\;\;\rho }^{\prime \ast A\sigma \gamma |}K_{\lambda \mu
\nu \alpha |\sigma \gamma }^{B}C_{\beta }^{C}=\gamma b_{2}+\partial _{\sigma
}\overset{(2)}{w}^{\sigma },  \label{colt95}
\end{equation}%
where $b_{2}$ and $\overset{(2)}{w}^{\sigma }$ must fulfill the properties
\begin{equation}
\mathrm{agh}\left( b_{2}\right) =2=\mathrm{agh}\left( \overset{(2)}{w}%
^{\sigma }\right) ,\qquad \mathrm{pgh}\left( b_{2}\right) =2,\qquad \mathrm{%
pgh}\left( \overset{(2)}{w}^{\sigma }\right) =3.  \label{colt96}
\end{equation}%
The above requirement takes place if and only if
\begin{equation}
c_{ABC}=0,  \label{colt97}
\end{equation}%
because the left-hand side of relation (\ref{colt95}) contains only
non-trivial elements of $H^{3}\left( \gamma \right) $ with the antighost
number equal to two, where the role of invariant polynomials is played by%
\begin{equation*}
-2c_{ABC}\varepsilon ^{\rho \lambda \mu \nu \alpha \beta }\mathcal{G}%
_{\;\;\;\;\;\;\;\;\;\;\rho }^{\prime \ast A\sigma \gamma |}K_{\lambda \mu
\nu \alpha |\sigma \gamma }^{B},
\end{equation*}%
which implies automatically $b_{2}=0$, and, on the other hand, this
expression cannot be written in a divergence-like form, such that we must
set $\overset{(2)}{w}^{\sigma }=0$. But $b_{2}=0$ and $\overset{(2)}{w}%
^{\sigma }=0$ simultaneously in (\ref{colt95}) lead to (\ref{colt97}), and
in consequence to%
\begin{equation}
a_{4}^{\mathrm{t}}=0.  \label{colt97a}
\end{equation}%
In conclusion, under the hypothesis that the maximum derivative order of the
interacting Lagrangian is equal to two, the first-order deformation can only
stop at antighost numbers $I\leq 3$.

In the case $I=3$ we have that%
\begin{equation}
a^{\mathrm{t}}=a_{0}^{\mathrm{t}}+a_{1}^{\mathrm{t}}+a_{2}^{\mathrm{t}%
}+a_{3}^{\mathrm{t}},  \label{colt98}
\end{equation}%
with $\gamma a_{3}^{\mathrm{t}}=0$, such that we can write (see (\ref{colt81}%
))
\begin{equation}
a_{3}^{\mathrm{t}}=\alpha _{3}\omega ^{3}\left( \mathcal{F}_{\lambda \mu \nu
\alpha }^{A},C_{\nu }^{A}\right) .  \label{colt99}
\end{equation}%
The consistency of $a^{\mathrm{t}}$ at antighost number two (the existence
of $a_{2}^{\mathrm{t}}$ as solution to the equation $\delta a_{3}^{\mathrm{t}%
}+\gamma a_{2}^{\mathrm{t}}=\partial _{\mu }\overset{(2)}{m}_{\mathrm{t}%
}^{\mu }$) requires that $\alpha _{3}d^{D}x$ is a non-trivial element from $%
H_{3}^{\mathrm{inv}D}\left( \delta |d\right) $. Because the elements with
the pure ghost number equal to three of a basis in the space of polynomials
in $\mathcal{F}_{\lambda \mu \nu \alpha }^{A}$ and $C_{\nu }^{A}$ are
spanned by%
\begin{equation}
\omega ^{3}:\left( C_{\beta }^{B},\mathcal{F}_{\lambda \mu \nu \alpha }^{B}%
\mathcal{F}_{\lambda _{1}\mu _{1}\nu _{1}\alpha _{1}}^{C}\mathcal{F}%
_{\lambda _{2}\mu _{2}\nu _{2}\alpha _{2}}^{D}\right) ,  \label{colt100}
\end{equation}%
and the general, non-trivial representatives of $H_{3}^{\mathrm{inv}D}\left(
\delta |d\right) $ are generated by the antifields $G^{\prime \ast A\nu
\alpha }$ (see (\ref{coltabledelta}) for $I=3$), we infer
\begin{equation}
a_{3}^{\mathrm{t}}=G_{\rho \sigma }^{\prime \ast A}\left( f_{1ABCD}^{\rho
\sigma \lambda \mu \nu \alpha \lambda _{1}\mu _{1}\nu _{1}\alpha _{1}\lambda
_{2}\mu _{2}\nu _{2}\alpha _{2}}\mathcal{F}_{\lambda \mu \nu \alpha }^{B}%
\mathcal{F}_{\lambda _{1}\mu _{1}\nu _{1}\alpha _{1}}^{C}\mathcal{F}%
_{\lambda _{2}\mu _{2}\nu _{2}\alpha _{2}}^{D}+f_{2AB}^{\rho \sigma \beta
}C_{\beta }^{B}\right) ,  \label{colt101}
\end{equation}%
where the coefficients denoted by $f$ must be some non-derivative, constant
tensors. The condition that the maximum derivative order of the interacting
Lagrangian is equal to two imposes the restrictions%
\begin{equation}
f_{1ABCD}^{\rho \sigma \lambda \mu \nu \alpha \lambda _{1}\mu _{1}\nu
_{1}\alpha _{1}\lambda _{2}\mu _{2}\nu _{2}\alpha _{2}}=0,  \label{colt102}
\end{equation}%
since otherwise the corresponding interacting term from $a_{0}^{\mathrm{t}}$
would be of order four in the space-time derivatives of the fields, and
hence we get
\begin{equation}
a_{3}^{\mathrm{t}}=f_{2AB}^{\rho \sigma \beta }G_{\rho \sigma }^{\prime \ast
A}C_{\beta }^{B}.  \label{colt103}
\end{equation}%
Asking now that $a_{3}^{\mathrm{t}}$ is a Lorentz covariant and Poincar\'{e}
invariant element defined on a space-time of dimension $D\geq 5$ leaves us
with the trivial solution
\begin{equation}
f_{2AB}^{\rho \sigma \beta }=0,  \label{colt104}
\end{equation}%
which further implies
\begin{equation}
a_{3}^{\mathrm{t}}=0.  \label{colt105}
\end{equation}%
In conclusion, the first-order deformation cannot stop in a non-trivial
manner also at the value $I=3$ of the antighost number.

Next, we pass to the situation where the non-integrated density of the
first-order deformation stops at antighost number two%
\begin{equation}
a^{\mathrm{t}}=a_{0}^{\mathrm{t}}+a_{1}^{\mathrm{t}}+a_{2}^{\mathrm{t}},
\label{colt106}
\end{equation}%
where $\gamma a_{2}^{\mathrm{t}}=0$, and hence, in agreement with (\ref%
{colt81}), we have that%
\begin{equation}
a_{2}^{\mathrm{t}}=\alpha _{2}\omega ^{2}\left( \mathcal{F}_{\lambda \mu \nu
\alpha }^{A}\right) .  \label{colt107}
\end{equation}%
(The ghosts $C_{\nu }^{A}$ no longer appear in $\omega ^{2}$ since their
pure ghost number is equal to three, while $\mathrm{pgh}\left( \omega
^{2}\right) =2$). We recall that a necessary condition for the existence of (%
\ref{colt107}) in antighost number one (the existence of $a_{1}^{\mathrm{t}}$
as solution to the equation $\delta a_{2}^{\mathrm{t}}+\gamma a_{1}^{\mathrm{%
t}}=\partial _{\mu }\overset{(1)}{m}_{\mathrm{t}}^{\mu }$) is that $\alpha
_{2}d^{D}x$ belongs to $H_{2}^{\mathrm{inv}D}\left( \delta |d\right) $. The
elements of pure ghost number equal to two of a basis in the space of
polynomials in $\mathcal{F}_{\lambda \mu \nu \alpha }^{A}$ are spanned by%
\begin{equation}
\omega ^{2}:\left( \mathcal{F}_{\lambda \mu \nu \alpha }^{A}\mathcal{F}%
_{\lambda _{1}\mu _{1}\nu _{1}\alpha _{1}}^{B}\right) ,  \label{colt108}
\end{equation}%
and the general, non-trivial representatives of $H_{2}^{\mathrm{inv}D}\left(
\delta |d\right) $ are built from the antifields $\mathcal{G}^{\prime \ast
A\mu \nu |\alpha }$ (see (\ref{coltabledelta}) for $I=2$), such that
\begin{equation}
a_{2}^{\mathrm{t}}=f_{1ABC}^{\rho \sigma \beta \lambda \mu \nu \alpha
\lambda _{1}\mu _{1}\nu _{1}\alpha _{1}}\mathcal{G}_{\rho \sigma |\beta
}^{\prime \ast A}\mathcal{F}_{\lambda \mu \nu \alpha }^{B}\mathcal{F}%
_{\lambda _{1}\mu _{1}\nu _{1}\alpha _{1}}^{C},  \label{colt109}
\end{equation}%
where the coefficients denoted by $f$ must be some non-derivative, constant
tensors. The derivative order hypothesis $a_{0}^{\mathrm{t}}$ requires
\begin{equation}
f_{1ABC}^{\rho \sigma \beta \lambda \mu \nu \alpha \lambda _{1}\mu _{1}\nu
_{1}\alpha _{1}}=0,  \label{colt110}
\end{equation}%
since otherwise, if consistent, component (\ref{colt109}) would lead to an $%
a_{0}^{\mathrm{t}}$ with three space-time derivatives acting on the fields.
Condition (\ref{colt110}) further implies%
\begin{equation}
a_{2}^{\mathrm{t}}=0,  \label{colt111}
\end{equation}%
and hence we can take $I\leq 1$ in the first-order deformation. The result (%
\ref{colt111}) emphasizes that the original, Abelian gauge algebra is rigid
with respect to the deformation procedure (since the existence of
non-trivial terms in $a_{2}^{\mathrm{t}}$ that are simultaneously linear in
the antifields with the antighost number equal to two and quadratic in
combinations of ghosts with the pure ghost number equal to one is not
allowed in the first-order deformation), such that the resulting
selfinteractions among the fields with the mixed symmetry $\left( 3,1\right)
$ might modify at most the original gauge transformations or the free
Lagrangian.

For $I=1$ the first-order deformation becomes%
\begin{equation}
a^{\mathrm{t}}=a_{0}^{\mathrm{t}}+a_{1}^{\mathrm{t}},  \label{colt112}
\end{equation}%
where the last component ($\gamma a_{1}^{\mathrm{t}}=0$) takes the generic
form (see (\ref{colt81}))
\begin{equation}
a_{1}^{\mathrm{t}}=\alpha _{1}\left( \left[ t_{A}^{\ast \lambda \mu \nu
|\alpha }\right] ,\left[ K_{\lambda \mu \nu \xi |\alpha \beta }^{A}\right]
\right) \omega ^{1}\left( \mathcal{F}_{\lambda \mu \nu \alpha }^{A}\right) .
\label{colt113}
\end{equation}%
The invariant polynomial $\alpha _{1}$ is linear in the antifields $%
t_{A}^{\ast \lambda \mu \nu |\alpha }$ and their derivatives (up to a finite
order) since these are the only objects of antighost number equal to one
from the BRST algebra, while
\begin{equation}
\omega ^{1}:\left( \mathcal{F}_{\lambda \mu \nu \alpha }^{B}\right) .
\label{colt116}
\end{equation}%
We mentioned in the above (see the end of Section \ref{col5}) that a
necessary condition for the consistency of $a^{\mathrm{t}}$ is that $\alpha
_{1}d^{D}x$ is a non-trivial element of $H_{1}^{D}\left( \delta |d\right) $,
which is infinite-dimensional. The impossible mission of computing $%
H_{1}^{D}\left( \delta |d\right) $ can be avoided if we demand from the
start the hypothesis on $a_{0}^{\mathrm{t}}$ to be of maximum derivative
order equal to two. This assumption is particularly useful at this stage
since it forbids the invariant polynomial $\alpha _{1}$ to depend on the
curvature tensors $K_{\lambda \mu \nu \xi |\alpha \beta }^{A}$ or their
space-time derivatives. Indeed, assuming that $\alpha _{1}$ effectively
depends on the curvature tensors, it follows that the component from (\ref%
{colt113}) with the minimum number of derivatives will be linear in the
undifferentiated antifields $t_{A}^{\ast \lambda \mu \nu |\alpha }$, in the
undifferentiated curvature tensors, as well as in the elements (\ref{colt116}%
), so it already contains three space-time derivatives. If consistent, it
would produce an $a_{0}^{\mathrm{t}}$ of order four in the space-time
derivatives of the fields. Therefore, we forbid the dependence on the
curvature tensors and remain with%
\begin{equation}
a_{1}^{\mathrm{t}}=\alpha _{1A}^{\mathrm{lin}}\left( \left[ t_{B}^{\ast
\lambda \mu \nu |\alpha }\right] \right) \mathcal{F}_{\lambda \mu \nu \alpha
}^{A}.  \label{colt117}
\end{equation}%
Moreover, the invariant polynomial $\alpha _{1}^{\mathrm{lin}}$ is further
restricted not to depend on the derivatives of $t_{B}^{\ast \lambda \mu \nu
|\alpha }$. This is because one can always move the derivatives (by making
an integration by parts) such as to act on $\mathcal{F}_{\lambda \mu \nu
\alpha }^{A}$, which provides purely trivial ($\gamma $-exact) contributions
to $a_{1}^{\mathrm{t}}$ (see (\ref{colt74})), which can be eliminated from
the first-order deformation.

The previous discussion allows us to state that the only eligible candidate
to $a_{1}^{\mathrm{t}}$ is defined in $D=6$ and reads%
\begin{equation}
a_{1}^{\mathrm{t}}\equiv a_{1}^{\mathrm{t}(D=6)}=c_{AB}\sigma ^{\alpha \beta
}\varepsilon ^{\lambda \mu \nu \lambda ^{\prime }\mu ^{\prime }\nu ^{\prime
}}t_{\lambda \mu \nu |\alpha }^{\ast A}\mathcal{F}_{\lambda ^{\prime }\mu
^{\prime }\nu ^{\prime }\beta }^{B}.  \label{colt117a}
\end{equation}%
Let us investigate the solutions in antighost number zero
\begin{equation}
\delta a_{1}^{\mathrm{t}(D=6)}+\gamma a_{0}^{\mathrm{t}(D=6)}=\partial _{\mu
}\overset{(0)}{m}_{\mathrm{t}}^{\mu }.  \label{colt121i}
\end{equation}%
In order to evaluate $\delta a_{1}^{\mathrm{t}(D=6)}$, we use the identity%
\begin{equation}
\varepsilon ^{\lambda \mu \nu \lambda ^{\prime }\mu ^{\prime }\nu ^{\prime
}}\sigma ^{\alpha \beta }\overset{(1)}{T}_{\lambda \mu \nu |\alpha }^{A}%
\overset{(2)}{T}_{\lambda ^{\prime }\mu ^{\prime }\nu ^{\prime }\beta }^{B}=-%
\frac{3}{4}\varepsilon ^{\lambda \mu \nu \lambda ^{\prime }\mu ^{\prime }\nu
^{\prime }}\sigma ^{\alpha \beta }\overset{(1)}{T}_{\lambda \mu \beta
|\alpha }^{A}\overset{(2)}{T}_{\nu \lambda ^{\prime }\mu ^{\prime }\nu
^{\prime }}^{B}  \label{colt121j}
\end{equation}%
(that takes place for any tensor $\overset{(1)}{T}_{\lambda \mu \nu |\alpha
}^{A}$ completely antisymmetric in its first three indices and for any
completely antisymmetric tensor $\overset{(2)}{T}_{\lambda ^{\prime }\mu
^{\prime }\nu ^{\prime }\beta }^{B}$) together with the first definition
from (\ref{colt55}). After some computation, we obtain that%
\begin{eqnarray}
&&\delta a_{1}^{\mathrm{t}(D=6)}=\gamma \left[ \frac{c_{AB}}{2}\left(
4-D\right) \varepsilon ^{\lambda \mu \lambda ^{\prime }\mu ^{\prime }\nu
^{\prime }\rho ^{\prime }}t_{\lambda \mu \left( \rho |\alpha \right)
}^{A}\partial _{\lambda ^{\prime }}\left( \sigma ^{\alpha \rho }\partial
^{\beta }t_{\mu ^{\prime }\nu ^{\prime }\rho ^{\prime }|\beta }^{B}-\partial
^{\alpha }t_{\mu ^{\prime }\nu ^{\prime }\rho ^{\prime
}|}^{B\;\;\;\;\;\;\;\;\rho }\right) \right]  \notag \\
&&+\partial _{\rho }j^{\rho }-\frac{c_{AB}}{2}\left( 4-D\right) \varepsilon
^{\lambda \mu \lambda ^{\prime }\mu ^{\prime }\nu ^{\prime }\rho ^{\prime }}%
\mathcal{T}_{\lambda \mu \left( \rho |\alpha \right) }^{A}\partial _{\lambda
^{\prime }}\left( \sigma ^{\alpha \rho }\partial ^{\beta }t_{\mu ^{\prime
}\nu ^{\prime }\rho ^{\prime }|\beta }^{B}-\partial ^{\alpha }t_{\mu
^{\prime }\nu ^{\prime }\rho ^{\prime }|}^{B\;\;\;\;\;\;\rho }\right) ,
\label{colt121k}
\end{eqnarray}%
where
\begin{equation}
\mathcal{T}_{\lambda \mu \nu |\alpha }^{A}\equiv 4\partial _{\left[ \lambda
\right. }\eta _{\left. \mu \nu \right] \alpha }^{A}+\partial _{\left[
\lambda \right. }\mathcal{G}_{\left. \mu \nu \right] |\alpha }^{A},\qquad
\mathcal{T}_{\lambda \mu \nu |\alpha }^{A}=3\mathcal{F}_{\lambda \mu \nu
\alpha }^{A}+\gamma t_{\lambda \mu \nu |\alpha }^{A}.  \label{colt67}
\end{equation}%
Comparing (\ref{colt121k}) with (\ref{colt121i}), we observe that the
existence of $a_{0}^{\mathrm{t}(D=6)}$ requires that the last terms from the
right-hand side of (\ref{colt121k}) either vanish or reduce to a full
divergence. It is clear from (\ref{colt67}) that they cannot reduce to a
divergence, and therefore must be set equal to zero, which further implies
\begin{equation}
c_{AB}=0,  \label{colt121l}
\end{equation}%
such that
\begin{equation}
a_{1}^{\mathrm{t}(D=6)}=0.  \label{colt121m}
\end{equation}%
Until now we showed that
\begin{equation}
a_{1}^{\mathrm{t}}=0,  \label{colt121}
\end{equation}%
and hence the first-order deformation may contain at most terms of antighost
number zero ($I=0$). The terms of antighost number one present in the
solution to the master equation are known to control the gauge symmetries,
such that (\ref{colt121}) expresses the fact that there are no consistent
selfinteractions in a collection of tensor fields $t_{\lambda \mu \nu
|\alpha }^{A}$ that deform the original gauge transformations, given in (\ref%
{colt7}).

In this manner, we are left with a sole possibility, namely that the
first-order deformation reduces to the deformed Lagrangian at order one in
the coupling constant%
\begin{equation}
a^{\mathrm{t}}=a_{0}^{\mathrm{t}}\left( \left[ t_{\lambda \mu \nu |\alpha
}^{A}\right] \right) ,  \label{colt122}
\end{equation}%
and thus it is subject to the equation
\begin{equation}
\gamma a_{0}^{\mathrm{t}}=\partial _{\mu }\overset{(0)}{m}_{\mathrm{t}}^{\mu
}.  \label{colt123}
\end{equation}%
Proceeding along a line similar to that employed in \cite{t31jhep}, it can
be shown that the solution to (\ref{colt123}) is purely trivial%
\begin{equation}
a_{0}^{\mathrm{t}}\left( \left[ t_{\lambda \mu \nu |\alpha }^{A}\right]
\right) =0.  \label{colt134}
\end{equation}%
Assembling the results expressed by (\ref{colt97a}), (\ref{colt105}), (\ref%
{colt111}), (\ref{colt121}), and (\ref{colt134}), we can state that%
\begin{equation}
S_{1}=0,  \label{colt135}
\end{equation}%
such that we can also take
\begin{equation}
S_{k}=0,\qquad k>1.  \label{colt136}
\end{equation}

Relations (\ref{colt135})--(\ref{colt136}) emphasize the following main
result of our paper:\textit{\ under the hypotheses of analyticity of
deformations in the coupling constant, space-time locality, Lorentz
covariance, Poincar\'{e} invariance, and conservation of the number of
derivatives on each field, there are no consistent selfinteractions in }$%
D\geq 5$\textit{\ for a collection of massless tensor fields with the mixed
symmetry }$\left( 3,1\right) $\textit{. }In other words, the presence of the
collection brings nothing new if compared to the case of a single tensor
field $t_{\lambda \mu \nu |\alpha }$.

\section{Selfinteractions for a collection of massless tensor fields with
the mixed symmetry $(2,2)$}

\subsection{Free model: Lagrangian formulation and BRST symmetry}

The starting point is given by the Lagrangian action for a finite collection
of free, massless tensor fields with the mixed symmetry of the Riemann
tensor in $D\geq 5$
\begin{eqnarray}
S_{0}^{\mathrm{r}}\left[ r_{\mu \nu |\alpha \beta }^{a}\right] &=&\int
\left( \frac{1}{8}\left( \partial ^{\lambda }r_{a}^{\mu \nu |\alpha \beta
}\right) \left( \partial _{\lambda }r_{\mu \nu |\alpha \beta }^{a}\right) -%
\frac{1}{2}\left( \partial _{\mu }r_{a}^{\mu \nu |\alpha \beta }\right)
\left( \partial ^{\lambda }r_{\lambda \nu |\alpha \beta }^{a}\right) \right.
\notag \\
&&-\left( \partial _{\mu }r_{a}^{\mu \nu |\alpha \beta }\right) \left(
\partial _{\beta }r_{\nu \alpha }^{a}\right) -\frac{1}{2}\left( \partial
^{\lambda }r_{a}^{\nu \beta }\right) \left( \partial _{\lambda }r_{\nu \beta
}^{a}\right) +\left( \partial _{\nu }r_{a}^{\nu \beta }\right) \left(
\partial ^{\lambda }r_{\lambda \beta }^{a}\right)  \notag \\
&&\left. -\frac{1}{2}\left( \partial _{\nu }r_{a}^{\nu \beta }\right) \left(
\partial _{\beta }r^{a}\right) +\frac{1}{8}\left( \partial ^{\lambda
}r_{a}\right) \left( \partial _{\lambda }r^{a}\right) \right) d^{D}x.
\label{colr1}
\end{eqnarray}%
Like in the previous section, we employ the flat Minkowski metric of `mostly
plus' signature $\sigma ^{\mu \nu }=\sigma _{\mu \nu }=(-++++\cdots )$. The
lowercase indices $a$, $b$, etc. stand for the collection indices and are
assumed to take discrete values $1$, $2$, $\ldots $, $n$. They are lowered
with a symmetric, constant and invertible matrix, of elements $k_{ab}$, and
are raised with the help of the elements $k^{ab}$ of its inverse. Each
tensor field $r_{\mu \nu |\alpha \beta }^{a}$ exhibits the mixed symmetry of
the Riemann tensor, so it is separately antisymmetric in the pairs $\left\{
\mu ,\nu \right\} $ and $\left\{ \alpha ,\beta \right\} $, is symmetric
under their permutation ($\left\{ \mu ,\nu \right\} \longleftrightarrow
\left\{ \alpha ,\beta \right\} $), and satisfies the identity $r_{\left[ \mu
\nu |\alpha \right] \beta }^{a}\equiv 0$. The notations $r_{\nu \beta }^{a}$
signify the traces of $r_{\mu \nu |\alpha \beta }^{a}$, $r_{\nu \beta
}^{a}=\sigma ^{\mu \alpha }r_{\mu \nu |\alpha \beta }^{a}$, which are
symmetric, $r_{\nu \beta }^{a}=r_{\beta \nu }^{a}$, while $r^{a}$ represent
their double traces , $r^{a}=\sigma ^{\nu \beta }r_{\nu \beta }^{a}$, which
are scalars. Action (\ref{colr1}) admits a generating set of gauge
transformations of the form%
\begin{equation}
\delta _{\xi }r_{\mu \nu |\alpha \beta }^{a}=\partial _{\mu }\xi _{\alpha
\beta |\nu }^{a}-\partial _{\nu }\xi _{\alpha \beta |\mu }^{a}+\partial
_{\alpha }\xi _{\mu \nu |\beta }^{a}-\partial _{\beta }\xi _{\mu \nu |\alpha
}^{a},  \label{colr8}
\end{equation}%
where the gauge parameters $\xi _{\mu \nu |\alpha }^{a}$ are bosonic
tensors, with the mixed symmetry $\left( 2,1\right) $. Just like in the case
of a single $\left( 2,2\right) $ field \cite{r22}, the gauge transformations
from (\ref{colr8}) are Abelian and off-shell, first-order reducible.
Consequently, the Cauchy order of this linear gauge theory is equal to three.

Related to the generators of the BRST algebra, the ghost spectrum contains
the fermionic ghosts $\mathcal{C}_{\alpha \beta |\mu }^{a}$ associated with
the gauge parameters and the bosonic ghosts for ghosts $\mathcal{C}_{\mu \nu
}^{a}$ corresponding to the first-order reducibility. Obviously, we will
require that $\mathcal{C}_{\alpha \beta |\mu }^{a}$ preserve the mixed
symmetry $(2,1)$ and the tensors $\mathcal{C}_{\mu \nu }^{a}$ remain
antisymmetric. The antifield spectrum comprises the antifields $r_{a}^{\ast
\mu \nu |\alpha \beta }$ associated with the original fields and those
corresponding to the ghosts, $\mathcal{C}_{a}^{\ast \mu \nu |\alpha }$ and $%
\mathcal{C}_{a}^{\ast \mu \nu }$. The antifields $r_{a}^{\ast \mu \nu
|\alpha \beta }$ still have the mixed symmetry $(2,2)$, $\mathcal{C}%
_{a}^{\ast \mu \nu |\alpha }$ the mixed symmetry $(2,1)$, and $\mathcal{C}%
_{a}^{\ast \mu \nu }$ are antisymmetric. Related to the traces of $%
r_{a}^{\ast \mu \nu |\alpha \beta }$, we will use the notations $r_{a}^{\ast
\nu \beta }=\sigma _{\mu \alpha }r_{a}^{\ast \mu \nu |\alpha \beta }$ and $%
r_{a}^{\ast }=\sigma _{\nu \beta }r_{a}^{\ast \nu \beta }$.

The BRST differential decomposes like in the previous section, as $s=\delta
+\gamma $, the corresponding degrees of the generators from the BRST complex
being valued like%
\begin{gather*}
\mathrm{pgh}\left( r_{\mu \nu |\alpha \beta }^{a}\right) =0,\quad \mathrm{pgh%
}\left( \mathcal{C}_{\mu \nu |\alpha }^{a}\right) =1,\quad \mathrm{pgh}%
\left( \mathcal{C}_{\mu \nu }^{a}\right) =2, \\
\mathrm{pgh}\left( r_{a}^{\ast \mu \nu |\alpha \beta }\right) =\mathrm{pgh}%
\left( \mathcal{C}_{a}^{\ast \mu \nu |\alpha }\right) =\mathrm{pgh}\left(
\mathcal{C}_{a}^{\ast \mu \nu }\right) =0, \\
\mathrm{agh}\left( r_{\mu \nu |\alpha \beta }^{a}\right) =\mathrm{agh}\left(
\mathcal{C}_{\mu \nu |\alpha }^{a}\right) =\mathrm{agh}\left( \mathcal{C}%
_{\mu \nu }^{a}\right) =0, \\
\mathrm{agh}\left( r_{a}^{\ast \mu \nu |\alpha \beta }\right) =1,\quad
\mathrm{agh}\left( \mathcal{C}_{a}^{\ast \mu \nu |\alpha }\right) =2,\quad
\mathrm{agh}\left( \mathcal{C}_{a}^{\ast \mu \nu }\right) =3.
\end{gather*}%
The actions of $\delta $ and $\gamma $ on the generators from the BRST
complex, which enforce the standard BRST properties, are given by
\begin{equation}
\gamma r_{\mu \nu |\alpha \beta }^{a}=\partial _{\mu }\mathcal{C}_{\alpha
\beta |\nu }^{a}-\partial _{\nu }\mathcal{C}_{\alpha \beta |\mu
}^{a}+\partial _{\alpha }\mathcal{C}_{\mu \nu |\beta }^{a}-\partial _{\beta }%
\mathcal{C}_{\mu \nu |\alpha }^{a},  \label{colr50}
\end{equation}%
\begin{equation}
\gamma \mathcal{C}_{\mu \nu |\alpha }^{a}=2\partial _{\alpha }\mathcal{C}%
_{\mu \nu }^{a}-\partial _{\left[ \mu \right. }\mathcal{C}_{\left. \nu %
\right] \alpha }^{a},\qquad \gamma \mathcal{C}_{\mu \nu }^{a}=0,
\label{colr51}
\end{equation}%
\begin{equation}
\gamma r_{a}^{\ast \mu \nu |\alpha \beta }=0,\qquad \gamma \mathcal{C}%
_{a}^{\ast \mu \nu |\alpha }=0,\qquad \gamma \mathcal{C}_{a}^{\ast \mu \nu
}=0,  \label{colr54}
\end{equation}%
\begin{equation}
\delta r_{\mu \nu |\alpha \beta }^{a}=0,\qquad \delta \mathcal{C}_{\mu \nu
|\alpha }^{a}=0,\qquad \delta \mathcal{C}_{\mu \nu }^{a}=0,  \label{colr52}
\end{equation}%
\begin{equation}
\delta r_{a}^{\ast \mu \nu |\alpha \beta }=\frac{1}{4}R_{a}^{\mu \nu |\alpha
\beta },\delta \mathcal{C}_{a}^{\ast \alpha \beta |\nu }=-4\partial _{\mu
}r_{a}^{\ast \mu \nu |\alpha \beta },\delta \mathcal{C}_{a}^{\ast \mu \nu
}=3\partial _{\alpha }\mathcal{C}_{a}^{\ast \mu \nu |\alpha }.
\label{colr53}
\end{equation}%
In the above $R_{a}^{\mu \nu |\alpha \beta }$ is defined by $\delta S_{0}^{%
\mathrm{r}}/\delta r_{a}^{\mu \nu |\alpha \beta }\equiv -\left( 1/4\right)
R_{\mu \nu |\alpha \beta }^{a}$ and reads as%
\begin{eqnarray}
&R_{\mu \nu |\alpha \beta }^{a}=&\Box r_{\mu \nu |\alpha \beta
}^{a}+\partial ^{\rho }\left( \partial _{\mu }r_{\alpha \beta |\nu \rho
}^{a}-\partial _{\nu }r_{\alpha \beta |\mu \rho }^{a}+\partial _{\alpha
}r_{\mu \nu |\beta \rho }^{a}-\partial _{\beta }r_{\mu \nu |\alpha \rho
}^{a}\right)  \notag \\
&&+\partial _{\mu }\partial _{\alpha }r_{\beta \nu }^{a}-\partial _{\mu
}\partial _{\beta }r_{\alpha \nu }^{a}-\partial _{\nu }\partial _{\alpha
}r_{\beta \mu }^{a}+\partial _{\nu }\partial _{\beta }r_{\alpha \mu }^{a}
\notag \\
&&-\frac{1}{2}\partial ^{\lambda }\partial ^{\rho }\left( \sigma _{\mu
\alpha }\left( r_{\lambda \beta |\nu \rho }^{a}+r_{\lambda \nu |\beta \rho
}^{a}\right) -\sigma _{\mu \beta }\left( r_{\lambda \alpha |\nu \rho
}^{a}+r_{\lambda \nu |\alpha \rho }^{a}\right) \right.  \notag \\
&&\left. -\sigma _{\nu \alpha }\left( r_{\lambda \beta |\mu \rho
}^{a}+r_{\lambda \mu |\beta \rho }^{a}\right) +\sigma _{\nu \beta }\left(
r_{\lambda \alpha |\mu \rho }^{a}+r_{\lambda \mu |\alpha \rho }^{a}\right)
\right)  \notag \\
&&-\Box \left( \sigma _{\mu \alpha }r_{\beta \nu }^{a}-\sigma _{\mu \beta
}r_{\alpha \nu }^{a}-\sigma _{\nu \alpha }r_{\beta \mu }^{a}+\sigma _{\nu
\beta }r_{\alpha \mu }^{a}\right)  \notag \\
&&+\partial ^{\rho }\left( \sigma _{\mu \alpha }\left( \partial _{\beta
}r_{\nu \rho }^{a}+\partial _{\nu }r_{\beta \rho }^{a}\right) -\sigma _{\mu
\beta }\left( \partial _{\alpha }r_{\nu \rho }^{a}+\partial _{\nu }r_{\alpha
\rho }^{a}\right) \right.  \notag \\
&&\left. -\sigma _{\nu \alpha }\left( \partial _{\beta }r_{\mu \rho
}^{a}+\partial _{\mu }r_{\beta \rho }^{a}\right) +\sigma _{\nu \beta }\left(
\partial _{\alpha }r_{\mu \rho }^{a}+\partial _{\mu }r_{\alpha \rho
}^{a}\right) \right)  \notag \\
&&-\frac{1}{2}\left( \sigma _{\mu \alpha }\partial _{\beta }\partial _{\nu
}-\sigma _{\mu \beta }\partial _{\alpha }\partial _{\nu }-\sigma _{\nu
\alpha }\partial _{\beta }\partial _{\mu }+\sigma _{\nu \beta }\partial
_{\alpha }\partial _{\mu }\right) r^{a}  \notag \\
&&-\left( \sigma _{\mu \alpha }\sigma _{\nu \beta }-\sigma _{\mu \beta
}\sigma _{\nu \alpha }\right) \left( \partial ^{\lambda }\partial ^{\rho
}r_{\lambda \rho }^{a}-\frac{1}{2}\Box r^{a}\right) .  \label{colr16}
\end{eqnarray}
The solution to the classical master equation for the free model under study
is given by
\begin{eqnarray}
S^{\mathrm{r}} &=&S_{0}^{\mathrm{r}}\left[ r_{\mu \nu |\alpha \beta }^{a}%
\right] +\int \left( r_{a}^{\ast \mu \nu |\alpha \beta }\left( \partial
_{\mu }\mathcal{C}_{\alpha \beta |\nu }^{a}-\partial _{\nu }\mathcal{C}%
_{\alpha \beta |\mu }^{a}+\partial _{\alpha }\mathcal{C}_{\mu \nu |\beta
}^{a}-\partial _{\beta }\mathcal{C}_{\mu \nu |\alpha }^{a}\right) \right.
\notag \\
&&\left. +\mathcal{C}_{a}^{\ast \mu \nu |\alpha }\left( 2\partial _{\alpha }%
\mathcal{C}_{\mu \nu }^{a}-\partial _{\left[ \mu \right. }\mathcal{C}%
_{\left. \nu \right] \alpha }^{a}\right) \right) d^{D}x.  \label{colr55}
\end{eqnarray}

\subsection{Computation of basic cohomologies}

In order to analyze the local equation satisfied by the non-integrated
density of the first-order deformation $a^{\mathrm{r}}$ ($S_{1}=\int a^{%
\mathrm{r}}d^{D}x$), written in local form and dual language, $sa^{\mathrm{r}%
}=\partial _{\mu }m_{\mathrm{r}}^{\mu }$, we proceed like in the previous
section. We ensure the space-time locality of the deformations by working in
the algebra of local differential forms with coefficients that are
polynomial functions in the fields, ghosts, antifields, and their space-time
derivatives (algebra of local forms). This means that we assume the
non-integrated density of the first-order deformation, $a^{\mathrm{r}}$, to
be a polynomial function in all these variables (algebra of local
functions). Next, we develop $a^{\mathrm{r}}$ according to the antighost
number and assume that this expansion contains a finite number of terms,
with the maximum value of the antighost number equal to $I$. Due to the
decomposition $s=\delta +\gamma $, this equation becomes equivalent to the
chain
\begin{eqnarray}
\gamma a_{I}^{\mathrm{r}} &=&\partial _{\mu }\overset{(I)}{m}_{\mathrm{r}%
}^{\mu },  \label{colr64} \\
\delta a_{I}^{\mathrm{r}}+\gamma a_{I-1}^{\mathrm{r}} &=&\partial _{\mu }%
\overset{(I-1)}{m}_{\mathrm{r}}^{\mu },  \label{colr65} \\
\delta a_{k}^{\mathrm{r}}+\gamma a_{k-1}^{\mathrm{r}} &=&\partial _{\mu }%
\overset{(k-1)}{m}_{\mathrm{r}}^{\mu },\qquad I-1\geq k\geq 1,
\label{colr66}
\end{eqnarray}%
where $\left( \overset{(k)}{m}_{\mathrm{r}}^{\mu }\right) _{k=\overline{0,I}%
} $ are some local currents, with $\mathrm{agh}\left( \overset{(k)}{m}_{%
\mathrm{r}}^{\mu }\right) =k$. Equation (\ref{colr64}) can be replaced in
strictly positive values of the antighost number (see \cite{r22th},
Corollary 3.1) with%
\begin{equation}
\gamma a_{I}^{\mathrm{r}}=0,\qquad I>0.  \label{colr67}
\end{equation}%
In conclusion, for $I>0$ we have that $a_{I}^{\mathrm{r}}\in H^{I}\left(
\gamma \right) $. We maintain the considerations from the previous section
on the uniqueness of $a_{I}^{\mathrm{r}}$ and $a^{\mathrm{r}}$.

Thus, in order to solve equations (\ref{colr67}) and (\ref{colr65})--(\ref%
{colr66}), it is necessary to compute the cohomology $H^{\ast }\left( \gamma
\right) $ in the algebra of local functions. Definitions (\ref{colr54}) and (%
\ref{colr50}) indicate that all the antifields%
\begin{equation}
\chi ^{\ast \Delta }=\left( r_{a}^{\ast \mu \nu |\alpha \beta },\mathcal{C}%
_{a}^{\ast \mu \nu |\alpha },\mathcal{C}_{a}^{\ast \mu \nu }\right) ,
\label{colr73ab}
\end{equation}%
the curvature tensors%
\begin{eqnarray}
F_{\mu \nu \lambda |\alpha \beta \gamma }^{a} &=&\partial _{\lambda
}\partial _{\gamma }r_{\mu \nu |\alpha \beta }^{a}+\partial _{\mu }\partial
_{\gamma }r_{\nu \lambda |\alpha \beta }^{a}+\partial _{\nu }\partial
_{\gamma }r_{\lambda \mu |\alpha \beta }^{a}  \notag \\
&&+\partial _{\lambda }\partial _{\alpha }r_{\mu \nu |\beta \gamma
}^{a}+\partial _{\mu }\partial _{\alpha }r_{\nu \lambda |\beta \gamma
}^{a}+\partial _{\nu }\partial _{\alpha }r_{\lambda \mu |\beta \gamma }^{a}
\notag \\
&&+\partial _{\lambda }\partial _{\beta }r_{\mu \nu |\gamma \alpha
}^{a}+\partial _{\mu }\partial _{\beta }r_{\nu \lambda |\gamma \alpha
}^{a}+\partial _{\nu }\partial _{\beta }r_{\lambda \mu |\gamma \alpha }^{a},
\label{colcurv}
\end{eqnarray}%
and all their space-time derivatives are non-trivial elements of $%
H^{0}\left( \gamma \right) $. The curvature tensors exhibit the mixed
symmetry $\left( 3,3\right) $. Simple computation shows that $H^{1}\left(
\gamma \right) =0$ and, moreover,
\begin{equation}
H^{2l+1}\left( \gamma \right) =0,\qquad l\geq 0.  \label{colhgzero}
\end{equation}%
By means of the last definition from (\ref{colr51}), we find that the ghosts
for ghosts, $\mathcal{C}_{\mu \nu }^{a}$, are non-trivial objects from $%
H^{\ast }\left( \gamma \right) $. Consequently, their space-time derivatives
are also $\gamma $-closed. From the first relation present in (\ref{colr51})
it follows that
\begin{equation}
\partial _{\left( \mu \right. }\mathcal{C}_{\left. \nu \right) \alpha
}^{a}\equiv \gamma \left( -\frac{1}{3}\mathcal{C}_{\alpha \left( \mu |\nu
\right) }^{a}\right) .  \label{colr77}
\end{equation}%
Formula (\ref{colr77}) emphasizes that the quantities $\partial _{\left( \mu
\right. }$$\mathcal{C}_{\left. \nu \right) \alpha }^{a}$ are trivial in $%
H^{\ast }\left( \gamma \right) $. Moreover, the objects $\partial _{\left[
\mu \right. }$$\mathcal{C}_{\left. \nu \alpha \right] }^{a}$ are not $\gamma
$-exact, and $\partial _{\left[ \mu \right. }$$\mathcal{C}_{\left. \nu %
\right] \alpha }^{a}$ (for $\mu ,\nu \neq \alpha $) belong to the same
equivalence class from $H^{\ast }(\gamma )$ like $\partial _{\left[ \mu
\right. }$$\mathcal{C}_{\left. \nu \alpha \right] }^{a}$, such that they
will also be non-trivial representatives of $H^{\ast }\left( \gamma \right) $%
. Meanwhile direct calculations produce the relations%
\begin{equation}
\partial _{\alpha }\partial _{\beta }\mathcal{C}_{\mu \nu }^{a}=\frac{1}{12}%
\gamma \left( 3\left( \partial _{\alpha }\mathcal{C}_{\mu \nu |\beta
}^{a}+\partial _{\beta }\mathcal{C}_{\mu \nu |\alpha }^{a}\right) +\partial
_{\left[ \mu \right. }\mathcal{C}_{\left. \nu \right] \,\left( \alpha |\beta
\right) }^{a}\right) ,  \label{colr78}
\end{equation}%
which show that all the space-time derivatives of the ghosts $\mathcal{C}%
_{\mu \nu }^{a}$ of order two or higher are trivial in $H^{\ast }\left(
\gamma \right) $. In conclusion, the only non-trivial combinations from $%
H^{\ast }\left( \gamma \right) $ built from the ghosts for ghosts are
polynomials in $\mathcal{C}_{\mu \nu }^{a}$ and $\partial _{\left[ \mu
\right. }$$\mathcal{C}_{\left. \nu \alpha \right] }^{a}$. Since $H^{0}\left(
\gamma \right) $ is non-trivial, so far we proved that only the
cohomological spaces $H^{2l}\left( \gamma \right) $, with $l\geq 0$, are
non-trivial. Therefore, equation (\ref{colr67}) possesses non-trivial
solutions only for even values of $I$, $I=2J$, where the general form of $%
a_{2J}^{\mathrm{r}}$ is given by
\begin{equation}
a_{I}^{\mathrm{r}}\equiv a_{2J}^{\mathrm{r}}=\alpha _{2J}\left( \left[ \chi
^{\ast \Delta }\right] ,\left[ F_{\mu \nu \lambda |\alpha \beta \gamma }^{a}%
\right] \right) e^{2J}\left( \mathcal{C}_{\mu \nu }^{a},\partial _{\left[
\mu \right. }\mathcal{C}_{\left. \nu \alpha \right] }^{a}\right) ,\qquad J>0.
\label{colr79}
\end{equation}%
Notation $\chi ^{\ast \Delta }$ follows from (\ref{colr73ab}). The
coefficients $\alpha _{I}([\chi ^{\ast \Delta }],[F_{\mu \nu \lambda |\alpha
\beta \gamma }^{a}])$ are nothing but the invariant polynomials (in form
degree zero) of the theory (\ref{colr1})--(\ref{colr8}).

Substituting solution (\ref{colr79}) in equation (\ref{colr65}) for $I=2J$
and taking into consideration definitions (\ref{colr50})--(\ref{colr51}), we
obtain that a necessary condition for the existence of non-trivial solutions
$a_{2J-1}^{\mathrm{r}}$ is that the invariant polynomials $\alpha _{2J}$
present in (\ref{colr79}) generate non-trivial elements from the
characteristic cohomology in antighost number $2J>0$ computed in the algebra
of local forms, $\alpha _{2J}d^{D}x\in H_{2J}^{D}\left( \delta |d\right) $.
As the free model under consideration is a linear gauge theory of Cauchy
order equal to three, the general results from the literature~\cite{gen1}
establish that%
\begin{equation}
H_{k}^{D}\left( \delta |d\right) =0,\qquad k>3.  \label{colr81}
\end{equation}%
In addition, it can be shown that if the invariant polynomial $\alpha _{k}$,
with $\mathrm{agh}\left( \alpha _{k}\right) =k\geq 3$, defines a trivial
element $\alpha _{k}d^{D}x\in H_{k}^{D}\left( \delta |d\right) $, then this
element can be taken to be trivial also in $H_{k}^{\mathrm{inv}D}\left(
\delta |d\right) $. The above results ensure that
\begin{equation}
H_{k}^{\mathrm{inv}D}\left( \delta |d\right) =0,\qquad k>3.  \label{colr81c}
\end{equation}%
Using definitions (\ref{colr53}), we find that the non-trivial, Poincar\'{e}%
-invariant representatives of $\left( H_{k}^{D}\left( \delta |d\right)
\right) _{k\geq 2}$ and $\left( H_{k}^{\mathrm{inv}D}\left( \delta |d\right)
\right) _{k\geq 2}$ are linearly generated by the following invariant
polynomials: for $k>3$ --- there are none; for $k=3$ --- $f_{\mu \nu }^{a}%
\mathcal{C}_{a}^{\ast \mu \nu }d^{D}x$; for $k=2$ --- $f_{\mu \nu \alpha
}^{a}\mathcal{C}_{a}^{\ast \mu \nu |\alpha }d^{D}x$. In the above the
coefficients denoted by $f$ stand for the components of some constant,
non-derivative tensors.

The previous results on $H_{I}^{D}\left( \delta |d\right) $ and $H_{I}^{%
\mathrm{inv}D}\left( \delta |d\right) $ allow us to eliminate successively
all the terms of antighost number strictly greater than two from the
non-integrated density of the first-order deformation. The last
representative is of the form (\ref{colr79}), where the invariant
polynomials necessarily define non-trivial elements from $H_{I}^{\mathrm{inv}%
D}\left( \delta |d\right) $ if $I=2$ or respectively from $H_{1}^{D}\left(
\delta |d\right) $ if $I=1$.

\subsection{Cohomological analysis of selfinteractions}

In order to develop the general method of construction of consistent
selfinteractions that can be added to the free action (\ref{colr1}), subject
to the gauge symmetry (\ref{colr8}), we initially solve equation (\ref{ec41}%
), responsible for the first-order deformation, and then approach its
consistency. We will work under the same hypotheses as before. The
derivative order assumption restricts the interaction Lagrangian to contain
only interaction vertices with maximum two space-time derivatives. Related
to the non-integrated density of the first-order deformation, we have seen
in the previous section that its component of highest antighost number, $I$,
is constrained to satisfy the relation $I=2J$ (see the result expressed by (%
\ref{colhgzero}) on $H^{\ast }(\gamma )$). On the other hand, results (\ref%
{colr81}) and (\ref{colr81c}) ensure that one can safely take $I\leq 2$.

In view of this, the first non-trivial situation is described by $I=2J=2>0$,
in which case we can write
\begin{equation}
a^{\mathrm{r}}=a_{0}^{\mathrm{r}}+a_{1}^{\mathrm{r}}+a_{2}^{\mathrm{r}},
\label{colr81a}
\end{equation}%
where $a_{2}^{\mathrm{r}}$ is the general, non-trivial solution to equation (%
\ref{colr67}), and hence, in agreement with formula (\ref{colr79}), has the
expression
\begin{equation}
a_{2}^{\mathrm{r}}=\alpha _{2}\left( \left[ \chi ^{\ast \Delta }\right] ,%
\left[ F_{\mu \nu \lambda |\alpha \beta \gamma }^{a}\right] \right)
e^{2}\left( \mathcal{C}_{\mu \nu }^{a},\partial _{\left[ \mu \right. }%
\mathcal{C}_{\left. \nu \alpha \right] }^{a}\right) .  \label{colr82}
\end{equation}%
The elements $e^{2}$ are spanned by $\mathcal{C}_{\mu \nu }^{a}$ and $%
\partial _{\left[ \mu \right. }$$\mathcal{C}_{\left. \nu \alpha \right]
}^{a} $, and $\alpha _{2}d^{D}x$ is a non-trivial element from $H_{2}^{%
\mathrm{inv}D}\left( \delta |d\right) $. Due to the fact that the general
representative of $H_{2}^{\mathrm{inv}D}\left( \delta |d\right) $ is linear
in the undifferentiated antifields $\mathcal{C}_{\mu \nu |\alpha }^{\ast a}$%
, we deduce that
\begin{equation}
a_{2}^{\mathrm{r}}=\mathcal{C}_{\mu \nu |\alpha }^{\ast a}\left( f_{ab}^{\mu
\nu \alpha \beta \gamma }\mathcal{C}_{\beta \gamma }^{b}+\bar{f}_{ab}^{\mu
\nu \alpha \beta \gamma \lambda }\partial _{\left[ \beta \right. }\mathcal{C}%
_{\left. \gamma \lambda \right] }^{b}\right) ,  \label{colr88}
\end{equation}%
where $f_{ab}^{\mu \nu \alpha \beta \gamma }$ and $\bar{f}_{ab}^{\mu \nu
\alpha \beta \gamma \lambda }$ are some non-derivative, constant tensors.
These constants cannot be simultaneously antisymmetric in the indices $%
\left\{ \mu ,\nu ,\alpha \right\} $ since the identity $\mathcal{C}%
_{a}^{\ast \left[ \mu \nu |\alpha \right] }\equiv 0$ would lead to $a_{2}^{%
\mathrm{r}}=0$. The last restriction (combined with the requirement $D\geq 5$%
) produces%
\begin{equation}
f_{ab}^{\mu \nu \alpha \beta \gamma }=0=\bar{f}_{ab}^{\mu \nu \alpha \beta
\gamma \lambda },  \label{colr92}
\end{equation}%
and hence%
\begin{equation}
a_{2}^{\mathrm{r}}=0,  \label{colr93a}
\end{equation}%
so the first-order deformation cannot end non-trivially at antighost number
two.

Due to the fact that the last representative $a_{I}^{\mathrm{r}}$ from the
first-order deformation is subject to the condition $I=2J$, we are left only
with the case $I=0$
\begin{equation}
a^{\mathrm{r}}=a_{0}^{\mathrm{r}}\left( \left[ r_{\mu \nu |\alpha \beta }^{a}%
\right] \right) ,  \label{colr94}
\end{equation}%
where $a_{0}^{\mathrm{r}}$ satisfies equation (\ref{colr64}) ($I=0$, so
equation (\ref{colr64}) is no longer equivalent to (\ref{colr67}))%
\begin{equation}
\gamma a_{0}^{\mathrm{r}}=\partial _{\mu }\overset{(0)}{m}_{\mathrm{r}}^{\mu
}.  \label{colr95}
\end{equation}%
Using a technique similar to that employed in \cite{r22}, we find that the
general solution to the last equation reduces to a linear combination of
double traces of the undifferentiated tensor fields $r_{\mu \nu |\alpha
\beta }^{a}$ (the analogue of the cosmological term for the Pauli--Fierz
Lagrangian)%
\begin{equation}
a_{0}^{\mathrm{r}}=\sum\limits_{a=1}^{n}c_{a}r^{a},  \label{colr97t}
\end{equation}%
with $c_{a}$ some real, arbitrary constants, such that
\begin{equation}
S_{1}=\sum\limits_{a=1}^{n}\int c_{a}r^{a}d^{D}x  \label{colr98}
\end{equation}%
represents the most general expression of the first-order deformation of the
solution to the master equation for a collection of massless tensor fields
with the mixed symmetry $\left( 2,2\right) $. Moreover, this solution is
already consistent to all orders in the coupling constant. Indeed, since $%
\left( S_{1},S_{1}\right) =0$, equation (\ref{ec42}) is satisfied with the
choice%
\begin{equation}
S_{2}=0,  \label{colr99}
\end{equation}%
and similarly, all the higher-order equations are fulfilled for
\begin{equation}
S_{3}=S_{4}=\ldots =0.  \label{colr100}
\end{equation}

Relations (\ref{colr98})--(\ref{colr100}) emphasize the following main
result of our paper:\textit{\ under the hypotheses of analyticity of
deformations in the coupling constant, space-time locality, Lorentz
covariance, Poincar\'{e} invariance, and conservation of the number of
derivatives on each field, there are no consistent selfinteractions in }$%
D\geq 5$\textit{\ for a collection of massless tensor fields with the mixed
symmetry of the Riemann tensor. The only terms that can be added to the free
Lagrangian action are given by a sum of cosmological terms, whose existence
does not modify the original gauge transformations.}

\section*{Acknowledgments}

One of the authors (E.M.B.) acknowledges financial support from the contract
464/2009 in the framework of the programme IDEI of C.N.C.S.I.S. (Romanian
National Council for Academic Scientific Research).

\end{document}